\documentclass[8pt]{article}
\usepackage[english]{babel}
\usepackage[utf8x]{inputenc}
\usepackage{amsmath}
\usepackage{mathrsfs}
\usepackage{graphicx}
\usepackage{mathtools}
\usepackage{amssymb}
\usepackage{subfigure}
\usepackage{datetime}
\newdate{date}{31}{07}{2019}
\date{\displaydate{date}}
\graphicspath{{Images/}}
\usepackage[colorinlistoftodos]{todonotes}
\usepackage{hyperref}

\hypersetup{
    colorlinks=true,
    linkcolor=blue,
    filecolor=magenta,      
    urlcolor=cyan,
}
 
\begin{document}
\begin{titlepage}
\newcommand{\HRule}{\rule{\linewidth}{0.1mm}} 
\center 
 
\textsc{\Large }\\[0.5cm] 
\textsc{\Large }\\[0.5cm] 
\includegraphics{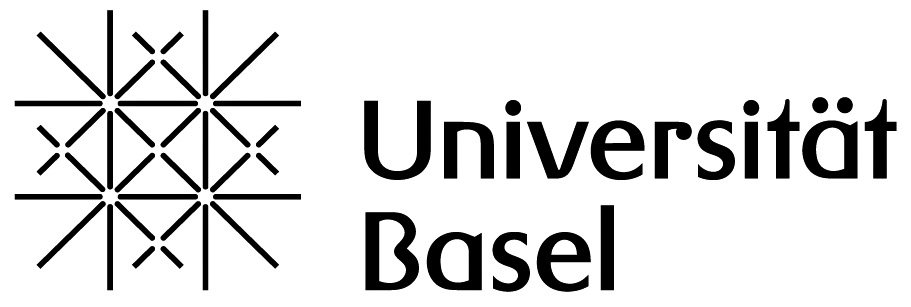}

\textbf{ Universit{\"a}t Basel, Department of Physical Chemistry, Quantum Technology (QuTe) group }
\vfill
\HRule \\[0.6cm]
{ \Large \bfseries Stabilization of $866$ nm laser with Pound-Drever-Hall (PDH) technique for quantum manipulation of Ca$^+$ ion in Paul trap}\\[0.1cm] 
\HRule \\[2cm]
 

\Large Summer internship report (2019)
\vfill

\begin{minipage}{0.4\textwidth}
\begin{flushleft} \large
\emph{Student:}\\
\textbf{Siddhant Singh}
\\{\textit{Indian Institute of Technology Kharagpur, 721302 India}}  
\end{flushleft}

\end{minipage}
\begin{minipage}{0.4\textwidth}
\begin{flushright} \large
\emph{Instructors:} \\
\textbf{Prof. Stefan Willitsch}\\ 
\textbf{Dr. Ziv Meir}
\\{\textit{Universit{\"a}t Basel, \quad Klingelbergstrasse 80, Switzerland}}
\end{flushright}
\end{minipage}\\[1cm]
{\large \displaydate{date}}\\[1cm] 
\vfill 

\end{titlepage}
\tableofcontents          
\listoffigures
\newpage


\section{Introduction}

Experimentalists seek and investigate well-isolated and controlled systems to explore the physical phenomenon. Ion trapping is one of the very well developed realization for the control of quantum systems. Ion trap manipulation of quantum states is done with the aid of lasers. The two most important qualities of a laser are its frequency and he linewidth about that frequency. A laser can be at the correct frequency, however, with huge linewidth. Or it can have a very narrow linewidth, however, drift over time. For intuitive reasons, these lasers must be as least noisy as possible. This noise usually comes from current fluctuations, mechanical disturbance, temperature fluctuations, electronic noise etc. \cite{ref:1} and tend to drift the laser frequency and or affect its linewidth over an interval of time. We require the laser to be nearly of the required transition frequency and also the least linewidth as possible. When we deal with lasers that do not have an inbuilt technology for stabilization of linewidth and frequency fluctuations, we need to develop external locking mechanisms for the laser. One such powerful method is the Pound-Drever-Hall technique \cite{ref:2}\cite{ref:3}. Laser locking or stabilization basically means we want to bring the laser back at its required point of operation whenever it drifts away. This can be tackled with a slow lock - this will only steer the laser frequency, however, will not affect the linewidth or a Fast lock - this will also narrow the laser linewidth. Going from slow to fast lock depends on the bandwidth of the lock. A simple locking can be done by reading the wavelength/frequency of the laser and then simply giving a digital feedback accordingly back to the laser whenever the numbers on the wavelength meter change. For this simple lock using the wavemeter, the bandwidth is very slow and only the laser frequency is steered. With PDH you can in principle to narrow the laser linewidth if we build a proper \textit{handle} on the laser. The basic idea is to lock the laser to a system which is much more stable than the laser itself in terms of frequency. It could be another much more stabilized laser or a system like an optical cavity. A different locking scheme, which was explored in this project, is to lock the laser to an ultra-low expansion optical cavity using the Pound-Drever-Hall (PDH) technique. This method allows for much tighter lock and also possible narrowing of the laser linewidth.  

In this article, I describe the theory relevant to the Pound-Drever-Hall technique and and its implementation for the stabilization of the 866 nm re-pumping (for Ca$^+$) laser in a main ion trap experiment.

\subsection{Manipulation of the Ca ion in an ion trap}
\begin{figure}[hbtp]
\centering
\includegraphics[scale=0.45]{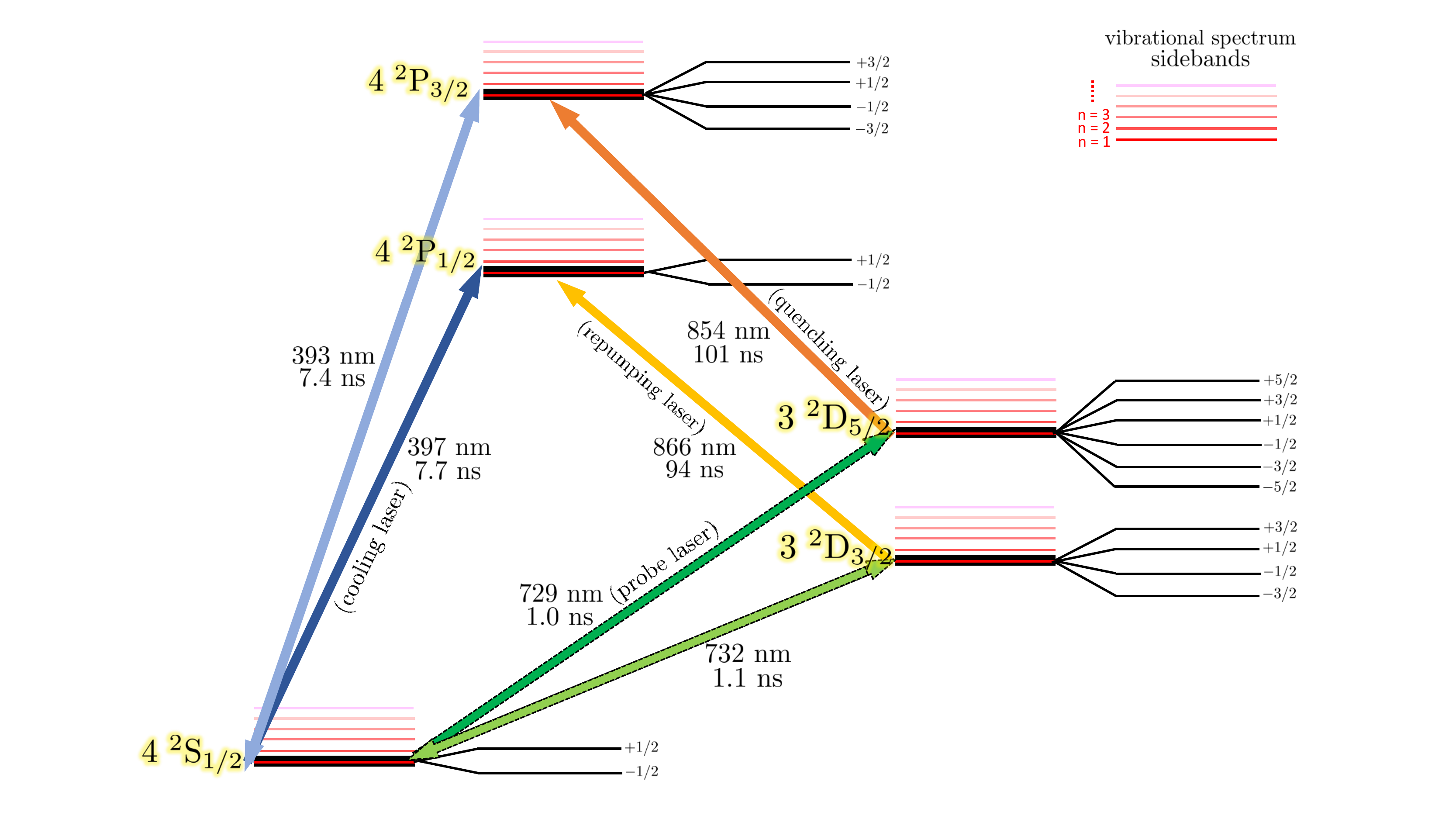}
\caption{The five lowest levels (in term symbols) of Ca$^+$ along with the Zeeman splittings, transition wavelengths and lifetimes. All transitions are dipole allowed transitions except the dotted outline ones.}
\label{fig:Ca+}
\end{figure}

Ion trap is a very well studied and experimentally realized technique \cite{ref:8}. It has very tremendous applications, such as, precision measurements, quantum state control \cite{ref:4}, mass spectroscopy \cite{ref:5} and recently emerging Trap Ion Quantum Computing \cite{ref:6} \cite{ref:7}. The two most common types of ion trap are the Penning trap, which forms a potential via a combination of electric and magnetic fields, and the Paul trap which forms a potential via a combination of static and oscillating electric fields. Ion trap realization of quantum information procession is a very promising model. Units of quantum information called qubits are stored in stable electronic states of each ion, and quantum information can be processed and transferred through the collective quantised motion of the ions, interacting by the Coulomb force (often via Paul trap). Lasers are applied to induce coupling between the qubit states (for single qubit operations) or between the internal qubit states and external motion states (for entanglement between qubits). For most of these applications, a well studied and well controlled ion is the Ca$^+$. Calcium belongs to the alkaline earth metals and has the electron configuration [Ar] 4s$^2$. The singly charged Ca$^+$ ion has a single electron outside the closed shells, and can for most purposes be considered as a one-electron atom. The energy levels of Ca$^+$ which are useful and exploited in ion trapping experiments are shown in figure \ref{fig:Ca+}.

After trapping, the ion is extremely hot with $\langle n \rangle  ∼ 10^7$, and many motional quanta must be removed; a high scattering rate is preferred. This can be obtained by cooling on the dipole allowed transition, S$_{1/2}$ ↔ P$_{1/2}$, which has a natural decay rate of $2\pi \cdot 20.7$ MHz. When the ion is in the P$_{1/2}$ state, it can decay to both S$_{1/2}$ and D$_{3/2}$. Decay to D$_{3/2}$ will interrupt the cooling as the lifetime of the D$_{3/2}$ level is long ($\sim 1$ s). This can be avoided by re-pumping on the D$_{3/2}$ ↔ P$_{1/2}$ transition. Given that decay to D$_{3/2}$ only occurs 1 out of 14.5 cycles and the momentum of the infrared photon is roughly half the momentum of the blue photon, it is expected that the final temperature is given by the Doppler limit on the blue transition. The final temperature is around  $T_D = 0.5$ mK. For trap frequency $\omega_z = 2\pi \cdot 500$ kHz, this temperature corresponds to $\langle n \rangle $ ∼ 20, and further cooling is needed to reach the ground state \cite{ref:4}.

The purpose of Doppler cooling is to remove a lot of kinetic energy. When initially an ion is trapped, it is most probably in its electronic ground state but in a much higher vibrational state (eigenstates of the ion trap oscillator). It has to be initialized in its ground state to begin with these techniques. The protocol to achieve the ground state cooling, to bring down the vibrational motion (sidebands) involves the manipulation of several of these electronic levels. 
Sideband cooling requires distinct motional sidebands. This can be realized on the narrow S$_{1/2}$ ↔ D$_{3/2}$ or S$_{1/2}$ ↔ D$_{5/2}$ quadrupole transitions with natural lifetimes of 1 s. The coupling strength on such transitions is significantly lower than on dipole allowed transitions and requires a significantly higher field strength. This is, however, readily available with commercial solid state laser systems or the diode lasers. Either transition is perfectly suited for sideband cooling, but the S$_{1/2}$ ↔ D$_{5/2}$ transition also provides the ability to easily discriminate between the two states. Due to the long life time of the D$_{5/2}$ level, a descent cooling rate requires broadening by coupling it to an auxiliary level. This can be achieved by driving the D$_{5/2}$ ↔ P$_{3/2}$ transition at 854 nm. If this transition is driven strongly, the sideband resolution is compromised. This can be overcome by using a low power or using interleaved pulses with excitation on the S$_{1/2}$ ↔ D$_{5/2}$ and subsequent repumping to the S$_{1/2}$ level via the P$_{3/2}$ level. When the ion is excited from S$_{1/2}$ to P$_{1/2}$, it has a finite probability to spontaneously decay to D$_{3/2}$ \cite{ref:5}. To restore this decay the re-pumping 866 nm laser is used to restore the excited quantum state.

Trapped ions are often detected by the light scattered during Doppler cooling. At saturation, a single ion can scatter many million photons every second, which makes it possible to detect the presence of the ion in short time. The fluorescence can also be used to detect the internal state of the ion by a technique called electron shelving \cite{ref:5}.

The idea behind this technique is to ‘shelve’ the \ electron in a state in which it does not fluoresce, e.g. the metastable state D$_{5/2}$ in $^{40}$Ca$^+$. If the ion is in the electronic ground state and the S$_{1/2}$ ↔ P$_{1/2}$ transition is driven, the ion will fluoresce from the scattered 397 nm photons. If the ion is in the D$_{5/2}$ state on the other hand, it will not fluoresce. The D$_{5/2}$ state has a lifetime of 1 s, which makes it possible to distinguish the states with high fidelity. On the narrow transition, it is possible to selectively transfer the ion depending on the motional state, and in this way the motional state can be mapped onto the internal state and read out.

\subsection{Laser stabilization and its necessity}

Frequency stabilization is needed for high precision because all lasers demonstrate frequency wander at some level. This instability is primarily due to temperature variations, mechanical imperfections, and laser gain dynamics, which change laser cavity lengths, laser driver current, voltage fluctuations, and many other factors. PDH locking offers one possible solution to this problem by actively tuning the laser to match the resonance condition of a stable reference cavity. PDH is an active locking mechanism. 

Probing dynamics on an atomic transition, like the $866$ nm $D_{1/2} - P_{1/2}$ transition in $^{40} $Ca$^+$, requires a laser with a narrow linewidth and absolute frequency stability. Typical grating stabilized diode lasers have linewidths of a few hundred kHz and can drift up to several MHz within a few minutes. This is too much to probe motional dynamics, and further stabilization is required. 
Methods for producing mode-locking in a laser may be classified as either 'active' or 'passive'. Active methods typically involve using an external signal to induce a modulation of the intracavity light. Passive methods do not use an external signal, but rely on placing some element into the laser cavity which causes self-modulation of the light.

The ultimate linewidth obtained from PDH stabilization depends on a number of factors. From a signal analysis perspective, the noise on the locking signal can not be any lower than that posed by the shot noise limit \cite{ref:1}. However, this constraint dictates how closely the laser can be made to follow the cavity. For tight locking conditions, the linewidth depends on the absolute stability of the cavity, which can reach the limits imposed by thermal noise. Using the PDH technique, optical linewidths below 40 mHz can be achieved \cite{ref:9}.

 The field of interferometric gravitational wave detection (such as at LIGO \cite{ref:10}) depends critically on enhanced sensitivity afforded by optical cavities. The PDH technique is also used when narrow spectroscopic probes of individual quantum states are required, such as atomic physics, time measurement standards, and quantum computers.

\section{Pound-Drever-Hall (PDH) theory}
\subsection{Overview}
The PDH stabilization relies on the use of an external optical cavity, which is its most important part. 
Since a wide range of conditions contribute to determine the linewidth produced by a laser, the PDH technique provides a means to control and decrease the laser's linewidth, provided an optical cavity that is more stable than the laser source. Alternatively, if a stable laser is available, the PDH technique can be used to stabilize and/or measure the instabilities in an optical cavity length \cite{ref:1}. We can also use the cavity with the PDH error signal to characterize the laser (if the laser is already stable this will measure the worst of the both - laser or cavity). The PDH technique responds to the frequency of laser emission independently of intensity, which is significant because many other methods that control laser frequency, such as a side-of-fringe lock are also affected by intensity instabilities.

Phase modulated light (an EOM can be used for this purpose), consisting of a carrier frequency and two side bands, is directed onto a two-mirror cavity. Light reflected off the cavity is measured using a fast response photodetector, the reflected signal consists of the two unaltered side bands along with a phase shifted carrier component. The photodetector signal is mixed down with an external local oscillator, which is in phase with the light modulation. After phase shifting and filtering, the resulting electronic signal gives a measure of how far the laser carrier is off resonance with the cavity and may be used as feedback for \textit{active} stabilization. The feedback is typically carried out using a \textit{PID Controller} which takes the PDH error signal readout and converts it into a voltage that can be fed back to the laser to keep it locked on resonance with the cavity \cite{ref:3}.

\subsection{The Optical Cavity (optical resonator)}
Optical cavity is simply an arrangement of two mirrors that can sustain standing waves for certain values of the wavelength which is dependent solely on the length of the cavity. It forms a standing wave cavity resonator for the electromagnetic waves. Light confined in a resonator will reflect multiple times from the mirrors, and due to the effects of interference, only certain patterns and frequencies of radiation will be sustained by the resonator, with the others being suppressed by destructive interference. Radiation patterns which are reproduced on every round-trip of the light through the resonator are the most stable, and these are the eigenmodes or simply the modes of the cavity resonator. Resonator modes are of two types: longitudinal modes, which differ in frequency from each other; and transverse modes, which may differ in both frequency and the intensity pattern of the light. The basic, or fundamental transverse mode of a resonator is a Gaussian beam.

One important point to note is the regime of the operation of the cavity. For PDH the cavity is operated in the high photon number regime or the classical regime unlike the low photon number or the quantum regime in treatments like the Cavity-QED where individual photonic modes can couple with other degrees of freedom. Here we only care about the beam transmission and reflection and we fire a trillion photons into the cavity! 

\subsubsection{Stability of the optical cavity}
\begin{figure}[hbtp]
\centering
\includegraphics[scale=0.5]{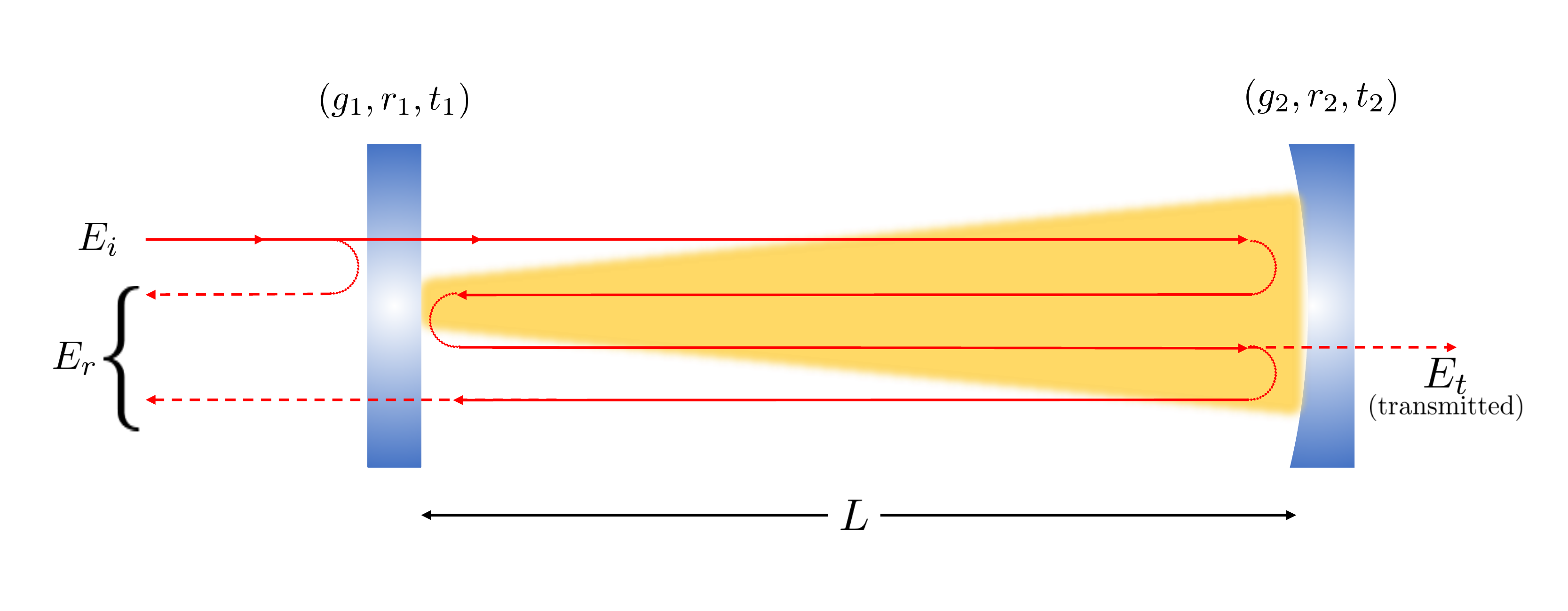}
\caption{An optical cavity of length $L$ with one plane mirror and other plano-concave mirror, with mirror parameters as $(g_i,r_i,t_i)$. All these parameters together completely define a cavity.}
\label{fig:cavitydep}
\end{figure}

The mirrors of the cavity could be spherical or plane or a \textit{sensible} combination of both type. The cavity I have used for my experiment is a plano-concave cavity as shown in figure \ref{fig:cavitydep}. Let the radius of curvature of the first mirror be $R_1$ and of second be $R_2$ and the length of the cavity $L$. Only certain ranges of values for $R_1$, $R_2$, and $L$ produce stable resonators in which periodic re-focussing of the intracavity beam is produced. If the cavity is unstable, the beam size will grow without limit, eventually growing larger than the size of the cavity mirrors and being lost \cite{ref:11}. This defines a stability domain for the geometry of the cavity as follows:
\begin{equation}
    0\leq  \underbrace{\left(1-\frac{L}{R_1}\right)}_{g_1}\underbrace{\left(1-\frac{L}{R_2}\right)}_{g_2} \leq 1
\end{equation}
$g_1$ and $g_2$ defines a hyperbolic region of stability for the resonator. This is a result of the \textit{Ray Transfer Matrix analysis} on the cavity \cite{ref:11}. Geometrically it means that a cavity is stable if the line segments between the mirrors and their centers of curvature overlap, but one does not lie entirely within the other. 

\subsubsection{Reflection and transmission in the cavity}
When light is incident on the entry mirror, a part of it is reflected and the rest leaks through and enters the resonator where it bounces back and forth between the front and back mirrors. The light is only resonant with the cavity if it interferes constructively after each round trip, that is, the optical length should match an integer number of wavelengths $(2L = N\lambda)$. Many equations in this section are well derived in the text \cite{ref:12}. This gives rise to a frequency dependent transmission profile with a set of equidistant peaks in the spectrum of wavelength/frequency. The separation between these peaks is called the \textit{free spectral range} given by
\begin{equation}
    \nu_{fsr}=c/2L
\end{equation}
where $c$ is the speed of light. Cavity \textit{finesse} which is a very important quantity is generally defined as 
\begin{equation}
    \mathscr{F}=\frac{\nu_{fsr}}{\nu_{1/2}}
    \label{eq:finlin}
\end{equation}
where $\nu_{1/2}$ is the linewidth of a transmission peak. The width of the peak depends upon the \textit{reflectivity} of the mirrors. High finesse means low linewidth (because the FSR is fixed for the length of the cavity) and a better cavity. In experiments, we tend to have a high finesse cavity. It is a measure for the average \textit{storage time} in the cavity. The higher the finesse, the larger the power circulating into the cavity and hence the more build up. It can be measured via the cavity ring down conversion (time needed to to observe an intensity drop of $1/e$ after suddenly switching off the injection light). Thus, the finesse depends only upon the reflection coefficients $r_i$ of the cavity and is more generally defined as
\begin{equation}
    \mathscr{F}=\pi \frac{\sqrt{r_1r_2l}}{1-r_1r_2l}
\end{equation}
where $l$ is the loss-coefficient ($=1$ for ideal cavity) which relates the amplitude of the light lost to the surrounding due to the non-ideal nature of the cavity. Losses
in a resonator can be induced for example by scattering of light in the propagation in the medium (if not vacuum) or on the mirrors’ surface itself.
In the very low loss limit of the cavity the finesse can be expressed as
\begin{equation}
\mathscr{F}=\frac{\pi r}{1-r^2}    
\label{eq:finesse}
\end{equation}
where $r$ is the reflection coefficient of the mirrors, assumed to be same for both mirrors (which is generally the case). Reflectivity should not be confused with reflectance of the mirror, usually denoted with 'R', which means the percentage reflection of incident intensity. For thin reflecting materials, like the mirrors here, it follows that $R=|r|^2$ here. Reflectivity, in general, may be complex.

Because of the continuous reflection of light between the mirrors, there is a power build-up in the cavity resonator; the enhancing factor or the build-up factor $S$ of the cavity is defined as
\begin{equation}
S=\frac{t_1^2}{(1-r_1r_2l)^2}.
\end{equation}

Light is injected into the cavity from the backside of the first mirror, taking advantage of the finite transmission of the first mirror encounter. Let the reflection and transmission of the first and second mirror be $r_1,t_1$ and $r_2,t_2$ respectively. These coefficients carry all the information needed to describe the optics of the mirrors.

If the mirrors are ideal and no light is lost or absorbed, then for each mirrors the following relation holds
\begin{equation}
    r_i^2+t_i^2=1.
\end{equation}
Consider that the incident field is $E_i=E_0e^{i\omega t}$. There is change of phase $\pi$ upon every reflection that wave faces from the first mirror. The successive round trip reflections give an overall phase of $2\pi$ and hence are trivial. And the phase acquired with every round trip is 
\begin{equation}
    \Phi=2\pi \frac{2L}{\lambda}=\frac{\omega}{\nu_{fsr}}
\end{equation}
When a saturation has been attained with the pumping and the exiting (cavity loss) light, we can write the total reflected field $E_r$ obtained as a series
\begin{equation}
    E_r=E_i(-r_1+t_1r_2t_1e^{i\Phi}+t_1r_2r_1r_2t_1e^{i2\Phi} + ...)
\end{equation}

We can sum up this geometric series and the resultant $E_r$ is
\begin{equation}
    E_r=E_i\underbrace{\left( -r_1 +\frac{t_1^2r_2 e^{i\Phi}}{1-r_1r_2e^{i\Phi}} \right)}_{F(\omega)}
    \label{eq:E_r}
\end{equation}

The function $F(\omega)$ is enough to exploit the properties of the cavity and use it for PDH technique.
When $\Phi=n2\pi$ or when $\omega=m\nu_{fsr}$ for some integers $n$ and $m$, $F(\omega)=0$ which means that the cavity transmits perfectly all the incident light at frequency exactly equal to $\omega$. 

\begin{figure}[hbtp]
\centering
\begin{subfigure}
\centering
\includegraphics[scale=0.25]{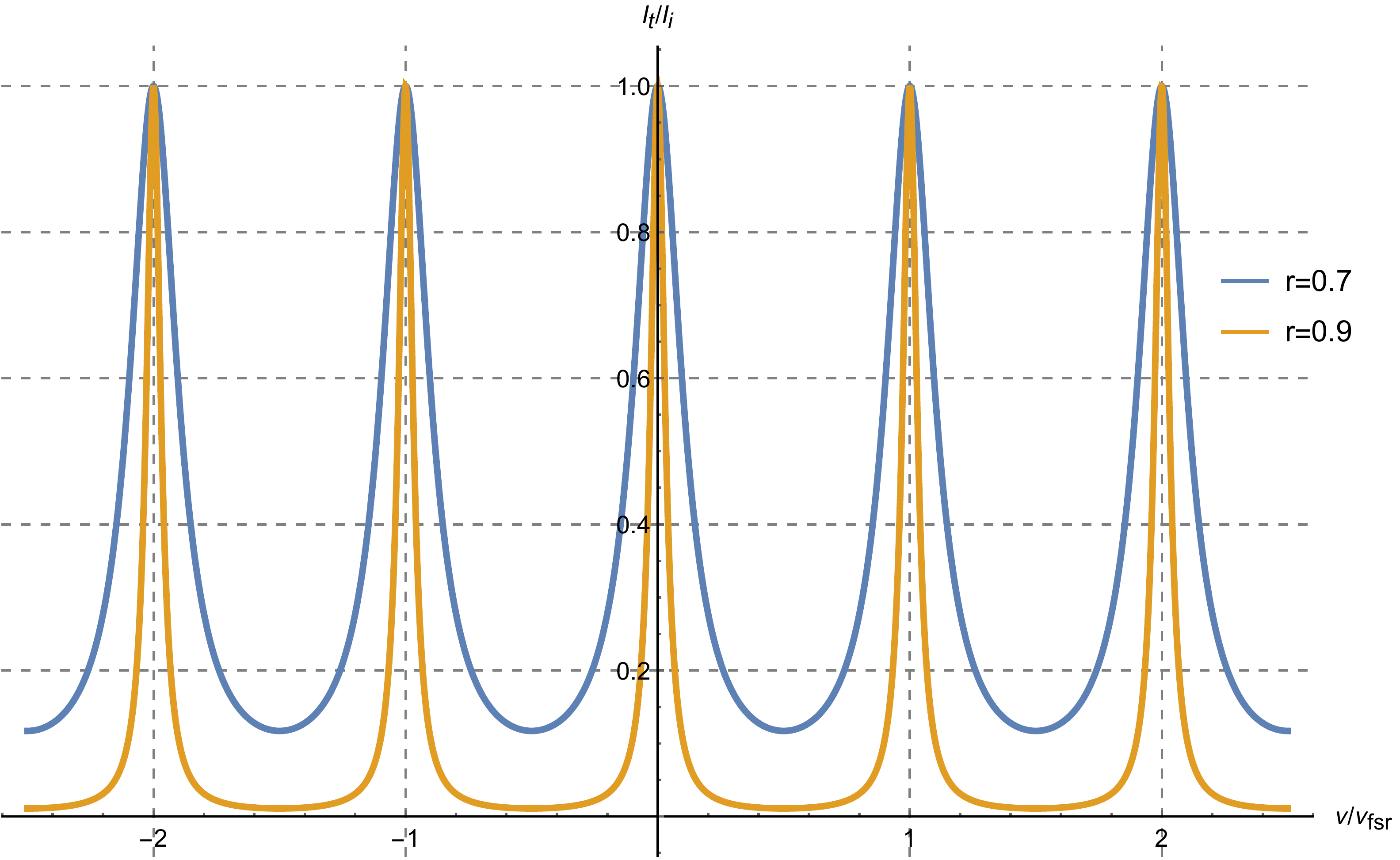}
\end{subfigure}
\begin{subfigure}
\centering
\includegraphics[scale=0.25]{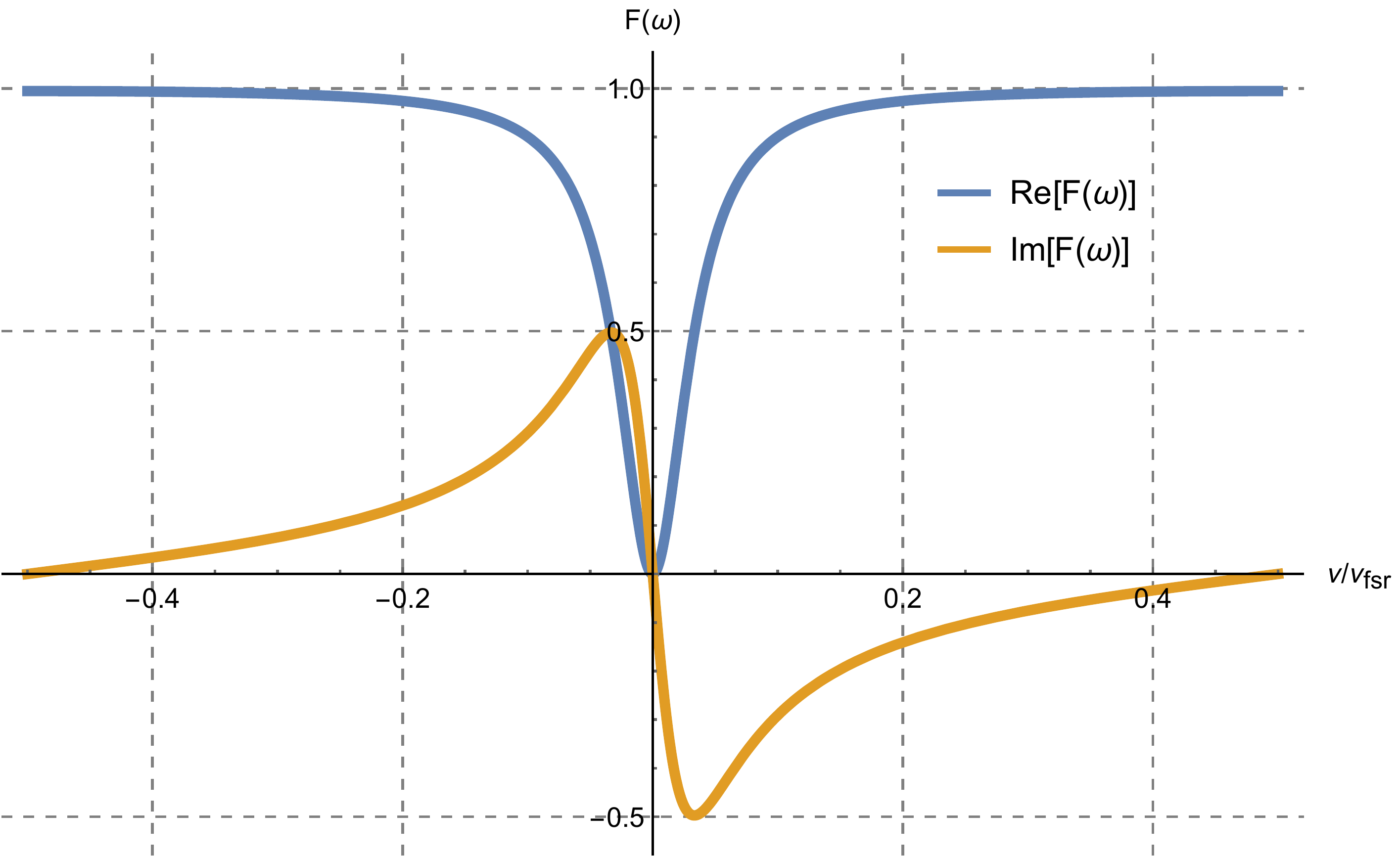}
\end{subfigure}
\caption{Cavity characteristics for transmission and the reflection: (a) shows the fractional transmitted intensity $I_t/I_i$ as a function of $\nu/\nu_{fsr}$ for different values of reflection coefficient ($r_1=r_2=r$, here). (b) shows the $\Re[F(\omega)]$ and $\Im[F(\omega)]$ near the resonance (for $r=0.9$ or $\mathscr{F}\approx 15$).}
\label{fig.cavitycharacterstics}
\end{figure}

When $r_1=r_2$, equation (\ref{eq:E_r}) can be re-expressed as 
\begin{equation}
   F(\omega)= r \frac{e^{i\omega/\nu_{fsr}}-1}{1-r^2e^{i\omega/\nu_{fsr}}}
\end{equation}
We plot the real and imaginary part of $F(\omega)$ in figure \ref{fig.cavitycharacterstics}. It is clear that more the value of $r$ close to unity, the more better the cavity is for frequency selectivity (the more sharp is the peak). From the plot, we can conclude that it is the imaginary part of $F(\omega)$ which carries the information about the direction of drift of frequency away from resonance. The principle of locking the frequency of the laser now means that one uses this information about which side has the laser frequency drifted over a time interval and by what amount, to generate a feedback signal to bring the laser back to the resonance with cavity.

\subsubsection{Transverse and longitudinal cavity modes}
For the optical cavity we are interested in the transverse cavity modes. While longitudinal cavity modes are just the standing wave patters formed due to interference effects of certain wavelength and their integer multiples. They are completely defined once we fix the laser frequency. However, there is an important point to note for the composite cavities. If the cavity is non-empty (i.e. contains one or more elements with different values of refractive index), the values of $L$ used are the optical path lengths for each element. The frequency spacing of longitudinal modes in the cavity is then given by:
\begin{equation}
    \Delta \nu_{fsr}=\frac{c}{2\sum_i n_iL_i }
\end{equation}
where $n_i$ is the refractive index of the $i$th element of length $L_i$. Both transverse and longitudinal waves may have longitudinal modes when confined to a cavity. For lasers with single transversal mode, the power per one longitudinal mode can be significantly increased by the coherent addition of lasers. Such addition allows one to both scale-up the output power of a single-transverse-mode laser and reduce number of longitudinal modes; because the system chooses automatically only the modes which are common for all the combined lasers. The reduction of the number of longitudinal modes determines the limits of the coherent addition. The ability to coherently add one additional laser is exhausted when one longitudinal mode, common for the combined lasers, lies within the spectral width of the gain; a subsequent addition will lead to loss of efficiency of the coherent combination and will not increase the power per longitudinal mode of such a laser. 

A transverse mode of electromagnetic radiation is a particular electromagnetic field pattern of the radiation in the plane perpendicular (i.e., transverse) to the radiation's propagation direction. Transverse modes occur because of boundary conditions imposed on the wave by the waveguide, like the optic fiber or the optical cavity.
Unguided electromagnetic waves in free space, or in a bulk isotropic dielectric, can be described as a superposition of plane waves; these can be described as TEM modes. However in any sort of waveguide where boundary conditions are imposed by a physical structure, a wave of a particular frequency can be described in terms of a transverse mode (or superposition of such modes). These modes generally follow different \textit{propagation constants} (describes the change in amplitude of the EM radiation after transmitting through a medium). When two or more modes have an identical propagation constant along the waveguide, then there is more than one modal decomposition possible in order to describe a wave with that propagation constant (for instance, a non-central Gaussian laser mode can be equivalently described as a superposition of Hermite-Gaussian modes or Laguerre-Gaussian modes).
Waveguide modes are classified as
\begin{itemize}
    \item \textbf{Transverse electromagnetic (TEM) modes} - Neither electric nor magnetic field in the direction of propagation.
    \item \textbf{Transverse electric (TE) modes} - No electric field in the direction of propagation.
    \item \textbf{Transverse magnetic (TM) modes} - No magnetic field in the direction of propagation. 
    \item \textbf{Hybrid modes} - Non-zero electric and magnetic fields in the direction of propagation.
\end{itemize}

The transverse profile of light circulating into a cavity must be an eigenfunction of the propagation equation (like a waveguide) of the light in the resonator. In a mathematical way, we can describe one round trip in a cavity using a propagation operator $K$. In this picture, the transverse field amplitude $E_{nm}'$ after one round trip is \cite{ref:11}
\begin{equation}
    E'_{nm}(x',y')=\int \int K(x',y',x,y)E_{nm}(x,y)dxdy=\gamma_{nm}E_{nm}(x,y)
\end{equation} 
where $E_{nm}(x,y)$ is the transverse field amplitude before the reflection. The above equation is an eigen-equation for $E'_{nm}$ with the eigenvalue $\gamma _{nm}$ (of the round-trip propagation operator).

\subsubsection{Laser cavity modes}
In a laser with cylindrical symmetry, the transverse mode patterns are described by a combination of a Gaussian beam profile (the base laser profile) with a Laguerre polynomial (imposed spatial boundary conditions). The modes are denoted TEM$_{pl}$ where $p$ and $l$ are integers labeling the radial and angular mode orders, respectively. The intensity at a point $(r,\varphi)$ (in polar coordinates) from the centre of the mode is given by \cite{ref:13}:
\begin{equation}
   I_{pl}(\rho,\varphi)= I_0 \rho^l \left[ L^l_p(\rho)\right]^2\cos (l\varphi)e^{-\rho} 
\end{equation}
where $\rho=2r^2/w^2$ and $L^l_p$ is the associated Laguerre polynomial of order $p$ and index $l$, and $w$ is the spot size of the mode corresponding to the Gaussian beam radius. 

\begin{figure}[hbtp]
\centering
\includegraphics[scale=0.22]{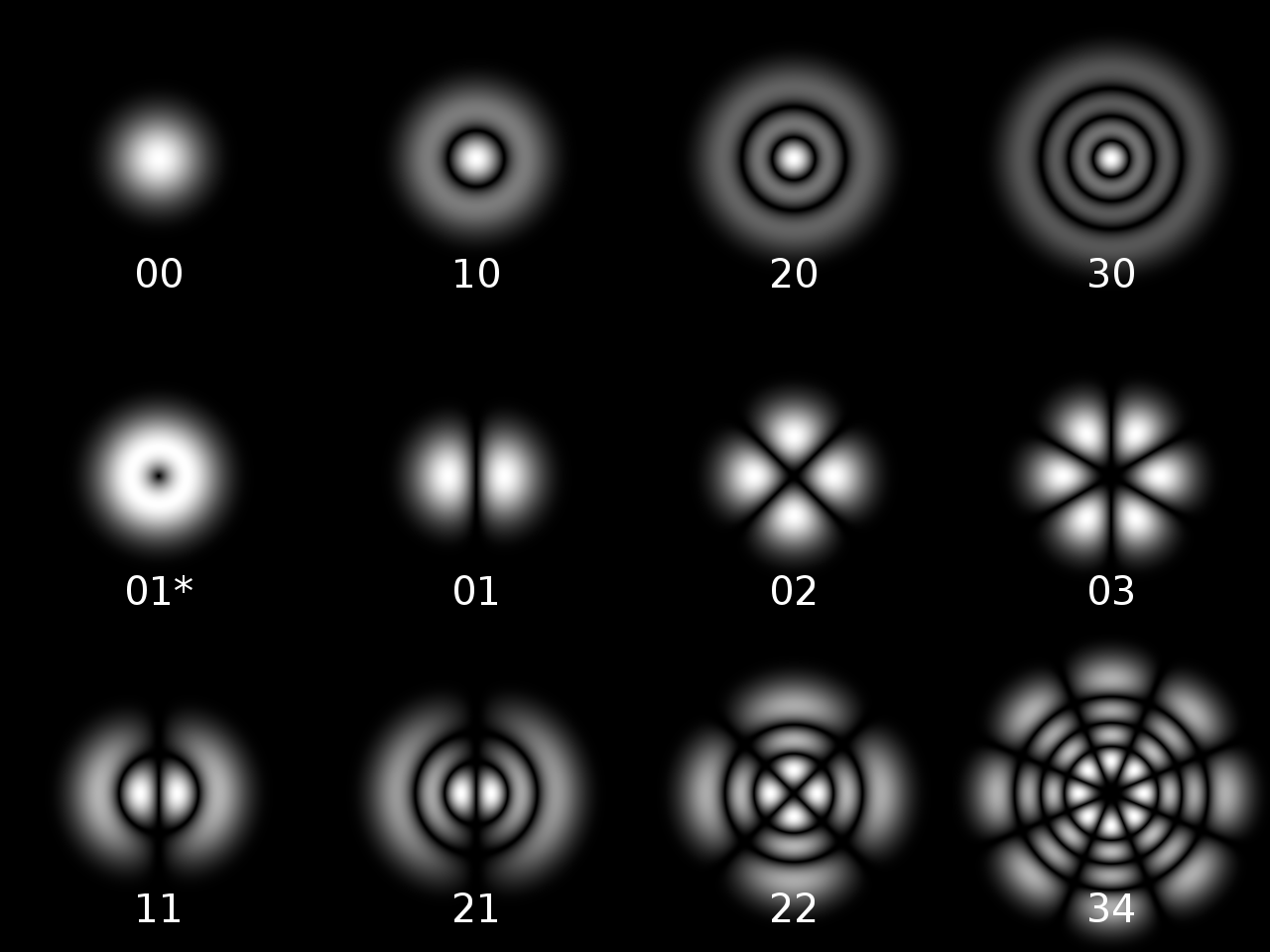}
\caption{Cylindrical transverse mode patterns TEM$_{[pl]}$. [\textit{Image credits: Wikimedia}]}
\label{fig:modes}
\end{figure}

With $ = l = 0$, the TEM$_{00}$ mode is the lowest order. It is the fundamental transverse mode of the laser resonator and has the same form as a Gaussian beam. This is the mode we want to lock our laser with. The pattern has a single lobe, and has a constant phase across the mode. Modes with increasing p show concentric rings of intensity, and modes with increasing $l$ show angularly distributed lobes. In general there are $2l(p+1)$ spots in the mode pattern (except for $l = 0$). In many lasers, the symmetry of the optical resonator is restricted by polarizing elements such as Brewster's angle windows. In these lasers, transverse modes with rectangular symmetry are formed. These modes are designated TEM$_{mn}$ with $m$ and $n$ being the horizontal and vertical orders of the pattern.  The corresponding intensity pattern is \cite{ref:13}
\begin{equation}
    I_{mn}(x,y,z)=I_0\left(\frac{w_0}{w}\right)^2\left[ H_m\left(\frac{\sqrt{2}x}{2}\right)\exp \left(\frac{-x^2}{w^2}\right)\right]^2\left[H_n\left(\frac{\sqrt{2}y}{w}\right)\exp \left(\frac{-y^2}{w^2}\right)\right]^2
\end{equation}
where $w_{0}$ and $w(z)$ are the waist, spot size for a Gaussian beam.$H_{k}$ is the $k^{th}$ Hermite polynomial. The TEM$_{00}$ mode corresponds to exactly the same fundamental mode as in the cylindrical geometry. Modes with increasing m and n show lobes appearing in the horizontal and vertical directions, with in general $(m + 1)(n + 1)$ lobes present in the pattern. The phase of each lobe of a TEM$_{mn}$ is offset by $\pi$ radians with respect to its horizontal or vertical neighbours. This is equivalent to the polarization of each lobe being flipped in direction. For my work (for only the laser locking part), the cylindrical symmetric modes are of relevance. In next chapter I will show how to obtain these laser modes and how they change with frequency, current temperature and other laser input parameters.

\subsubsection{Gaussian beam}
For laser beams, the Gaussian beam is a good approximation. Here, I will give a brief review of the Gaussian beams and their characteristics \cite{ref:11}. The Gaussian beam propagation is driven by the following equation, the first part represents a spherical wave propagating along $z$ with real radius of curvature $R(z)$, while the second represents the finite transverse amplitude variation which has a Gaussian form:
\begin{equation}
    u(x,y,z)=\frac{1}{q(z)}\exp\left[-ik\frac{x^2+y^2}{2R(z)}-\frac{x^2+y^2}{w^2(z)} \right]
\end{equation}
where $R(z)$ is the radius of curvature of the beam, $w(z)$ is the beam radius and $q(z)$ is
the complex radius defined as
\begin{equation}
    q(z)=\frac{1}{R(z)}-i\frac{\lambda}{\pi w^2(z)}
\end{equation}
An important feature of Gaussian beams is that it is possible to predict the entire propagation only knowing the beam waist $w_0$ (smallest transverse radius) and the wavelength of the light in the medium $\lambda$. In particular
\begin{equation}
    w(z)=w_0\sqrt{1+\left(\frac{z}{z_R}\right)^2}
\end{equation}
and 
\begin{equation}
    R(z)=z+\frac{z_R^2}{z}
\end{equation}
where $z_R$ is called Rayleigh length and is defined by
\begin{equation}
    z_R=\frac{\pi w_0^2}{\lambda}.
\end{equation}
If the radii of curvature of the two cavity mirrors match exactly the ones of the wavefronts, each mirror will perfectly reflect the Gaussian beam on itself. In this case, the system is a stable optical resonator. Generally, if the beam radii at the first mirror, second mirror and the centre of the cavity are $w_1, w_2$ and $w_0$ respectively, then it holds that \cite{ref:11}
\begin{equation}
    \begin{split}
        &w_1^2=\frac{L\lambda}{\pi}\sqrt{\frac{g_2}{g_1(1-g_1g_2)}}\\
        &w_2^2=\frac{L\lambda}{\pi}\sqrt{\frac{g_1}{g_2(1-g_1g_2)}}\\
        &w_0^2=\frac{L\lambda}{\pi}\sqrt{\frac{g_1g_2(1-g_1g_2)}{(g_1+g_2-2g_1g_2)^2}}.
    \end{split}
\end{equation}
With a beam centered on an aperture, the power $P$ passing through a circle of radius $r$ in the transverse plane at position $z$ is
\begin{equation}
    P(r,z)=P_0[1-e^{-2r^2/w^2(z)}]    
\end{equation}
where
\begin{equation}
    P_0=\pi I_0w_0^2/2.
\end{equation}

\subsection{Electro-Optic-Modulator}
In an Electro-Optic-Modulator (EOM), a signal-controlled element exhibiting an electro-optic effect is used to modulate a beam of light. The modulation may be imposed on the phase, frequency, amplitude, or polarization of the beam. Modulation bandwidths extending into the gigahertz range are possible with the use of laser-controlled modulators. 

The electro-optic effect is the change in the refractive index of a material resulting from the application of a DC or low-frequency electric field. This is caused by forces that distort the position, orientation, or shape of the molecules constituting the material. Generally, a nonlinear optical material (organic polymers have the fastest response rates, and thus are best for this application) with an incident static or low frequency optical field will see a modulation of its refractive index.

The simplest kind of EOM consists of a crystal, such as lithium niobate, whose refractive index is a function of the strength of the local electric field. That means that if lithium niobate is exposed to an electric field, light will travel more slowly through it. But the phase of the light leaving the crystal is directly proportional to the length of time it takes that light to pass through it. Therefore, the phase of the laser light exiting an EOM can be controlled by changing the electric field in the crystal.

Note that the electric field can be created by placing a parallel plate capacitor across the crystal. Since the field inside a parallel plate capacitor depends linearly on the potential, the index of refraction depends linearly on the field, and the phase depends linearly on the index of refraction, the phase modulation must depend linearly on the potential applied to the EOM.

For the PDH method, EOM is useful for generation of sidebands around a frequency value (as we will see later via an approximation). This is what tells us by the response of the cavity that it is drifting away from a desired value of the frequency.

\subsection{Pound-Drever-Hall technique}
The Pound-Drever-Hall technique relies on sideband modulation to extract the phase of the reflected carrier. Sidebands can be imposed on the light by modulating the phase of the light periodically. This can be realized for example with an Electro-Optic Modulator (EOM) where the refractive index depends on the applied electric field. 
\subsubsection{Generation of sidebands}
The unmodulated signal is $E_i=E_0e^{i\omega t}$ then we perform the Phase-Modulation of the wave by passing it through the EOM and driving the EOM at a desired resonant frequency. Phase modulation (PM) is a modulation pattern that encodes information as variations in the instantaneous phase of a carrier wave. The phase of a carrier signal is modulated to follow the changing voltage amplitude of modulation signal. The peak amplitude and frequency of the carrier signal remain constant, but as the amplitude of the information signal changes, the phase of the carrier changes correspondingly. The following treatment is also discussed in detail in \cite{ref:4}, \cite{ref:3}.

Now we apply a sinusoidally varying potential voltage to the EOM with frequency $\Omega$  and small amplitude $\beta$ . The resulting signal becomes
\begin{equation}
    E_i \xrightarrow{EOM} E_0e^{i(\omega t +\beta \sin \Omega t)}
    \label{eq:EOM}
\end{equation}
where $E_0$ is the electric field amplitude and $\omega$ the angular frequency of the light as before. $\beta$ is the so-called modulation index indicating the peak phase shift of the field. Our regime of interest is the low amplitude modulation $\beta \ll 1$ where we can treat the sidebands in first order only. Using the general integral representation of the Bessel functions of the first kind $J_\alpha$ for $\Re (x)> 0$
\begin{equation}
    J_\alpha (x)=\frac{1}{\pi}\int_0^\pi \left(\cos(\alpha \tau)-x \sin (\tau) \right)-\frac{\sin (\alpha \pi)}{\pi}\int _0^\infty e^{-x\sinh t-\alpha t}dt
\end{equation}

which for integer values of $\alpha = n$ takes the form
\begin{equation}
    J_n(x)=\frac{1}{\pi}\int_0^\pi \cos(n\tau x -\sin (\tau))d\tau=\frac{1}{2\pi}\int_{-\pi}^\pi e^{x \sin \tau-n\tau}d\tau.
\end{equation}

Expressing equation (\ref{eq:EOM}) in the form of the above integral representation of the Bessel function and making the small amplitude approximation and only keeping the terms up to the first order we have the following
\begin{equation}
    E_i\approx E_0 \left( J_0(\beta)e^{i\omega t}+J_1(\beta)e^{i(\omega+\Omega)t}-J_1(\beta)e^{i(\omega-\Omega)t} \right).
\end{equation}
which is an approximate version of the equation:
\begin{equation}
    E_i=Ae^{i\omega t}\left(J_0(\beta)+ \sum_{k=1}^\infty J_k(\beta)e^{ik\Omega t}+\sum_{k=1}^\infty (-1)^kJ_{k}(\beta)e^{-ik\Omega t}\right)
\end{equation}
We can also think of this as an small angle approximation
\begin{equation}
    e^{i(\omega t+\beta \sin (\Omega t))}\approx e^{i\omega t}(1+i\beta \sin (\Omega t))=e^{i\omega t}\left(1+\beta \frac{e^{i \Omega t}}{2}-\beta \frac{e^{-i\Omega t}}{2}\right)
\end{equation}
but as evident, the Bessel approximation is much more safe and general as it keeps the information for the amplitudes which are the first order functions of the modulation index $\beta$. When $\beta$ is not much smaller than unity, higher order sidebands must be included for a correct description of the field. Amplitude modulation and polarization modulation can also be used for other motives.

The above form shows the presence of a carrier with frequency $\omega$ and two first order sidebands with frequencies $\omega \pm \Omega$. 
If $P_{in}=|E_0|^2$ is the initial incident power of the laser beam then the carrier power is clearly $P_c=J_0^2(\beta)P_{in}$ and the first order sidebands is $P_s=J_1^2(\beta)P_{in}$. For the approximation of the small modulation depth the approximation
\begin{equation}
    P_c+2P_s \approx P_{in}
\end{equation}
holds very well. Basically we distribute some fraction of the initial incident power to the sidebands. The higher order sidebands are very negligible in their fractional power.

\subsubsection{Modulation beam reflection and the PDH error signal}
When this modulated field is incident on the resonator, the reflected field amplitude (here we assume that the beam incident on the cavity has three independent frequencies acting as three different beams) will be given by (using equation (\ref{eq:E_r})):
\begin{equation}
    E_r=E_0(F(\omega)J_0(\beta)e^{i\omega t}+F(\omega+\Omega)J_1(\beta)e^{i(\omega+\Omega)t}-F(\omega-\Omega)J_1(\beta)e^{i(\omega-\Omega)t})
\end{equation}
When this field is recorded by a detector, it records the intensity given by \cite{ref:3}\cite{ref:4}
\begin{equation}
    \begin{split}
        I_r=|E_r|^2=&|E_0|^2(|F(\omega)J_0(\beta)|^2+|F(\omega+\Omega)J_1(\beta)|^2+|F(\omega-\Omega)J_1(\beta)|^2 \\
        &+2J_0(\beta)J_1(\beta)\Re \left[F(\omega)F^*(\omega+\Omega)-F^*(\omega)F(\omega-\Omega)\right]\cos (\Omega t)\\
        &+2J_0(\beta)J_1(\beta)\Im \left[F(\omega)F^*(\omega+\Omega)-F^*(\omega)F(\omega-\Omega)\right]\sin (\Omega t)\\
        &+2J_1^2(\beta)\Re \left[F(\omega-\Omega)F^*(\omega+\Omega)\right]\cos(2\Omega t)\\
        &+2J_1^2(\beta)\Im \left[F(\omega-\Omega)F^*(\omega+\Omega)\right]\sin(2\Omega t).
    \end{split}
    \label{eq.PDH}
\end{equation}

The resulting signal is a wave with nominal frequency $\omega$ but with a beating pattern oscillating at two distinct frequencies $\Omega$ and $2\Omega$. We are interested in the two terms that are oscillating at the modulation frequency $\Omega$ because they sample the phase of the reflected carrier. There are two terms in this expression: a sine term and a cosine term. Usually, only one of them will be important. The other will vanish. Which one vanishes and which one survives depends on the modulation frequency. Define $\mathscr{I}(\omega)=F(\omega)F^*(\omega+\Omega)-F^*(\omega )F(\omega-\Omega)$. At low modulation frequencies (slow enough for the internal field of the cavity to have time to respond)  $\mathscr{I}(\omega)$ is purely real and only the cosine term survives. At high $\Omega$ near the cavity resonance, it is purely imaginary, and only the sine term is important. Without the cavity, the phase of the different terms match up to a pure phase modulation which is undetected by the photo diode. This is seen by setting all the $F(\omega)$-terms to unity, in which case the sine and cosine terms disappear i.e., the beating pattern vanishes. With the cavity, the phase is frequency dependent and this gives rise to a beating pattern when the light is \textit{slightly} (order of few kHz or MHz depending upon the cavity linewidth) detuned from cavity resonance. The important term is the $\cos(\Omega t)$ term, as this samples the phase of the carrier. This term can be isolated with a mixer and a low pass filter and is later used to generate the error signal for the feedback and locking. The $2\Omega$ terms will actually still be left, but this is a result from truncating the Bessel series and including higher order terms would make those terms cancel out too trivially. Also, the amplitude of these terms is still very small in comparison to the leading order ($J_1J_0$ and $J_1^2$ respectively) and furthermore if we keep these terms they will be further suppressed more by the action of the mixer and the low pass filter in contract to the leading order.

\subsubsection{Extraction of the error signal}
A mixer is an electronic device where the output is the product of the two inputs - in this case the reference oscillator $[\sin(\Omega t)]$ and
the PDH signal from equation (\ref{eq.PDH}). The PDH signal contains several harmonic frequencies, more generally
\begin{equation}
    \sum_i A_i(t)\sin (\Omega_i't+\phi_i)
\end{equation}
The output of the mixer gives
\begin{equation}
    \epsilon=\sin(\Omega t)\sum_i A_i(t)\sin(\Omega '_it+\phi_i)=\frac{1}{2}\sum_iA_i(t)\left[\cos((\Omega'_i-\Omega)t+\phi_i)-\cos((\Omega'_i+\Omega)t+\phi_i)\right].
\end{equation}
For an input with a single frequency of $\Omega′$, the output will be a sum of two frequencies $\Omega + \Omega'$ and $\Omega-\Omega'$. If $\Omega = \Omega′$, the term $\cos((\Omega − \Omega′)t + \phi)$ is a stationary signal which can be isolated with a low pass filter. The amplitude is explicitly time dependent $[A_i(t)]$, but as long as the amplitude varies slower than the cutoff of the low pass filter these variations will pass through; it is possible to extract the amplitudes of the signals oscillating at frequency $\Omega$. In equation (\ref{eq.PDH}) we have both a sine and a cosine wave oscillating at frequency $\Omega$. This can be incorporated in the phase $\phi$ and combining these equations we get:

\begin{equation}
    \epsilon=|E_0|^2J_0(\beta)J_1(\beta)\left(\Re[\mathscr{I}(\omega)]\sin \phi +\Im[\mathscr{I}(\omega)]\cos \phi\right)
\end{equation} 
Depending on $\phi$, the error signal, $\epsilon$, is a linear combination of the two illustrated signals. The signal for $\phi=0$ has a large slope and is close to linear just around the resonance; this is exactly what we want to determine the frequency of the light (this is where PDH works the best). Experimentally, $\phi$ is determined by the phase difference between the optical and electrical paths. $\phi=0$ is then realized by including an adjustable phase delay on the reference oscillator. When the carrier is near resonance and the modulation frequency is high enough that the sidebands are not, the sidebands will be totally reflected, $F(\omega\pm\Omega)=1$. Then the error signal becomes
\begin{equation}
    \epsilon=2|E_0|^2J_0(\beta)J_1(\beta)\Im[F(\omega)]
    \label{eq.errorper}
\end{equation}
specifically for $\phi=0$. The error signal samples the phase of the carrier! 

Close to a resonance, $\delta$ will be close to a multiple of $2\pi$, and we can write $\delta = 2\pi N + \Delta\omega/\nu_{fsr}$, where $N$ is an integer and $\Delta\omega$ is the detuning from resonance. For small $\Delta\omega$ we can expand $e^{i\Phi}$ from equation (\ref{eq:E_r}) in a power series, and if we truncate this after the first two terms we get:
\begin{equation}
    F(\Delta\omega)\approx r\frac{i\Delta\omega/\nu_{fsr}}{1-r^2}=\frac{\mathscr{F}}{\pi}\frac{i\Delta\omega}{\nu_{fsr}}=\frac{i}{\pi}\frac{\Delta\omega}{\nu_{1/2}}
\end{equation}
Combining this with equation (\ref{eq.errorper}) results in:
\begin{equation}
    \epsilon=2|E_0|^2J_0(\beta)J_1(\beta)\frac{\Delta\omega}{\pi \nu_{1/2}}
    \label{eq.errorsig}
\end{equation}
The amplitude of this error signal (with no modulation, the error signal is zero) depends on the modulation depth (through $J_1 (\beta)$), and the incident optical power. From equation (\ref{eq.errorsig}), we also see that the error signal is inversely proportional of the cavity linewidth, $\nu_{1/2}$. The narrower the the cavity linewidth the steeper the error signal we'll get.
 By realizing that the peak-to-peak amplitude $\epsilon_{pp}=2|E_0|^2J_0(\beta)J_1(\beta)$ we get a relation between the frequency deviation $(\Delta\nu = \Delta\omega/2\pi)$ and the error signal given by:
\begin{equation}
    \Delta\nu=\frac{\nu_{1/2}}{2\epsilon_{pp}}\epsilon
\end{equation}
This result is valid for small frequency detunings just around resonance and can be used to translate fluctuations in the error signal to fluctuations in the frequency of the laser, and thereby to evaluate the lock performance and estimate the linewidth of the laser.
 
\begin{figure}[hbtp]
\centering
\includegraphics[scale=0.4]{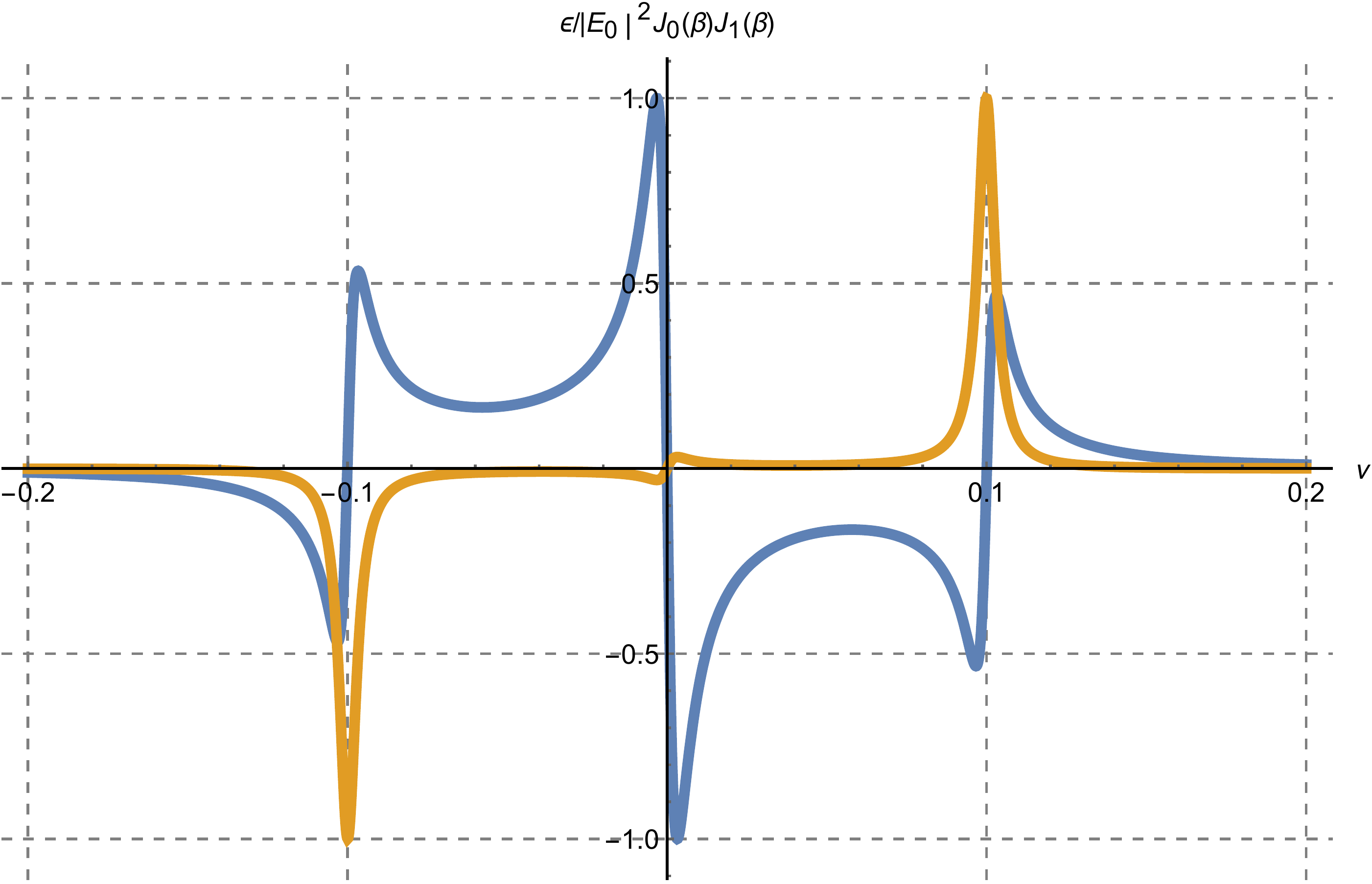}
\caption{PDH signal $\epsilon /|E_0|^2J_0(\beta)J_1(\beta)$ as a change of carrier frequency $\nu=\omega/2\pi$ with $\Omega=0.1$ units, $r=0.99$ and $\phi=0$ (Blue) $\&$ $\phi=\pi/2$ (Orange). The $\phi=0$ is the signal we want in this example. But it changes for different parameters which can be well chosen when obtaining the PDH signal in experiment and tuning with with the phase shifter.}
\label{fig:PDHsig}
\end{figure}
 
With the PDH technique, we can realize an antisymmetric signal with strong dependence on the detuning close to resonance; we have a good frequency reference for feeding back unwanted deviations this way. Feedback is the final step and can be well explained by the Control theory of feedback. Before the feedback, let me elaborate more on the modulation frequency regimes.
\begin{itemize}
    \item \textbf{Slow modulation:} The effective instantaneous frequency is the rate of change of the absolute phase \cite{ref:3},
    \begin{equation}
        \omega(t)=\frac{d}{dt}(\omega t+\beta \sin (\Omega t))=\omega +\beta \Omega \cos (\Omega t)
    \end{equation}
    and the reflected power is simply $P_{ref}=P_{in}|F(\omega)|^2$. To find the rate of change of the power, we can Taylor expand it around the carrier frequency as
    \begin{equation}
        P_{ref}(\omega+\beta \Omega \cos (\Omega t))\approx P_{ref}(\omega)+P_{in}\frac{d}{dt}|F(\omega)|^2\cdot \beta \Omega \cos (\Omega t).
    \end{equation}
    We dithered the frequency of the laser \textit{adiabatically}, slowly enough that the standing wave inside the cavity was always in equilibrium with the incident beam. We can express this in the quantitative model by making $\Omega$ very small. In this regime the expression
    $\mathscr{I}(\omega)$ becomes 
    \begin{equation}
        \mathscr{I}(\omega)\approx 2\Re \left[F(\omega) \frac{d}{d\omega}F^*(\omega)\right]=\frac{d}{dt}|F(\omega)|^2
    \end{equation}
    which is purely real! Then the only term that survives in equation (\ref{eq.PDH}) is the cosine term. In this operating regime we can make another useful approximation which holds, $\sqrt{P_cP_s}\approx P_{in}\beta /2$ then the reflected power is 
    \begin{equation}
        P_{ref}=(\text{constant terms})+P_{in}\frac{d}{d\omega}|F(\omega)|^2 \cdot \beta \Omega \cos(\Omega t) + (2\Omega\text{ terms}).
    \end{equation}
    The mixer will filter out everything but the term that varies as $\cos (\Omega t)$. We may have to adjust the phase of the signal before we feed it into the mixer (which can simply be done by introducing a delay in one of the signals into the mixer). The Pound–Drever–Hall error signal is then
    \begin{equation}
        \epsilon =P_{in}\frac{d}{d\omega}|F(\omega)|^2\cdot \beta \Omega \approx 2\sqrt{P_cP_s}\Omega \frac{d}{d\omega}|F(\omega)|^2.
    \end{equation}
    \item \textbf{Fast modulation frequency (best regime for PDH stabilization when near resonance)}: When the carrier is near resonance and the modulation frequency is high enough that the sidebands are not, we can assume that the sidebands are totally reflected, that is, $F(\omega \pm \Omega)\approx -1$. In this limit $\mathscr{I}(\omega)\approx -i 2 \Im[F(\omega)]$ (which is exactly what I described in the usual PDH case using another argument, however there are few more details to catch here) is purely imaginary. The cosine term now in equation (\ref{eq.PDH}) is negligible and the sine term is important for obvious reason. The error signal now takes the form
    \begin{equation}
        \epsilon=-2\sqrt{P_cP_s}\Im[\mathscr{I}(\omega)]
    \end{equation}
    This is identical to the previously obtained error signal in equation (\ref{eq.errorper}). Near resonance the reflected power essentially vanishes, since $|F(\omega)|^2\approx 0$. We do want to retain terms to first order in $F(\omega)$, however, to approximate the error signal, 
    \begin{equation}
        P_{ref}\approx 2P_s-4\sqrt{P_cP_s}\Im[F(\omega)]\sin (\Omega t)+ ...
    \end{equation}
    Using the same arguments as in the first treatment from equation (\ref{eq.errorper}) to equation (\ref{eq.errorsig}) we can write down 
    \begin{equation}
        \epsilon=-\frac{4}{\pi} \sqrt{P_cP_s}\frac{\delta \omega}{\delta \nu_{1/2}} 
    \end{equation}
    which is same as before (the minus sign is just a outcome of the convention, it will be settled automatically by the feedback and one does not need to worry about it). That the error signal is linear near resonance allows us to use the standard tools of control theory to suppress frequency noise. It can also be re-written as 
    \begin{equation}
        \epsilon=D \delta f
    \end{equation}
    where 
    \begin{equation}
        D=-8\frac{\sqrt{P_cP_s}}{\delta \nu_{1/2}}
    \end{equation}
    is called the frequency discriminant.
\end{itemize}
 
\subsection{Laser feedback and locking}
Now that the error signal has been obtained, one needs to feed it back to lock the laser. We can analyze how fast the system can respond to a change in input by applying a step signal at the input and monitor the signal at the output as a function of time. This response can be calculated through inverse Laplace transform. This section is extracted from \cite{ref:4}.

\subsubsection{Feedback loop}
\begin{figure}[hbtp]
\centering
\includegraphics[scale=0.3]{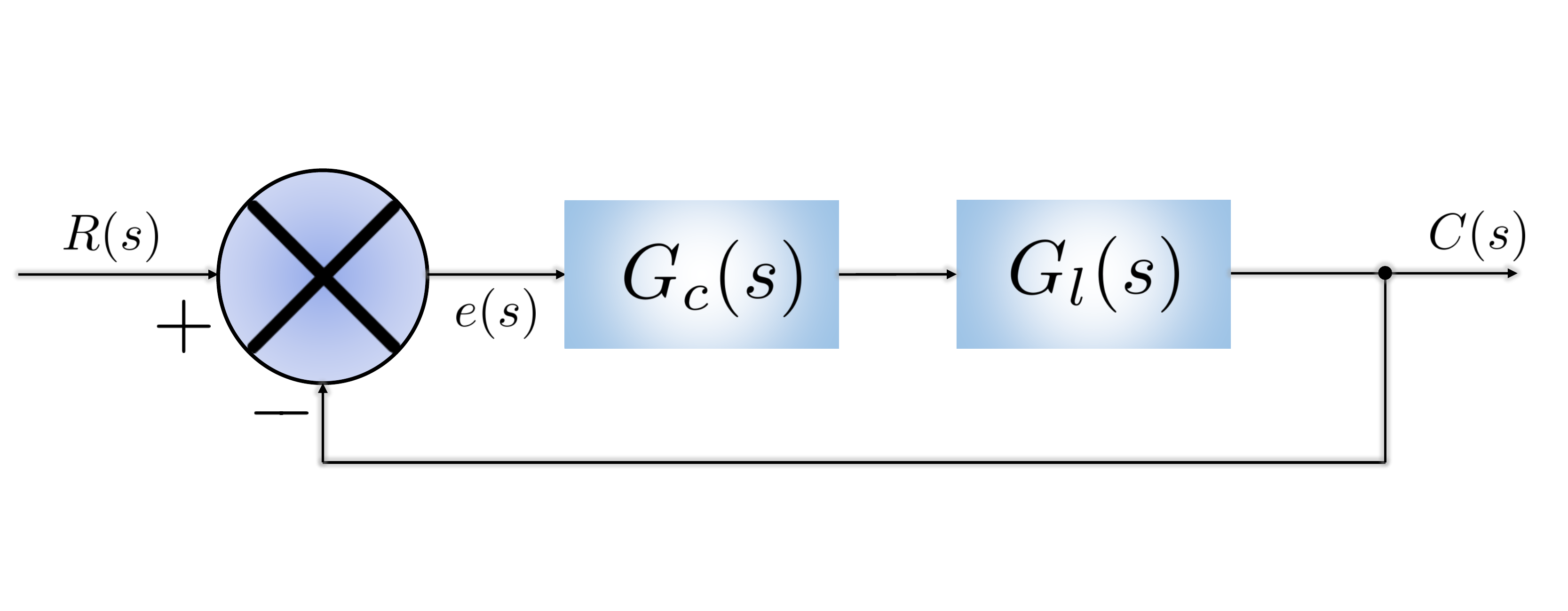}
\caption{Block diagram of the closed loop system with $G_c(s)$ representing the controller and $G_l(s)$ as the laser.}
\label{fig:closedloop}
\end{figure}

It is possible to improve the response of the circuit by placing it in a closed loop as illustrated in the block diagram in figure \ref{fig:closedloop}. In a closed loop, the output, $C(s)$, is subtracted from the input, $R(s)$, in the summing junction. The error, $e(s)$, is then sent through a controller with transfer function $G_c(s)$ and finally through the system we want to regulate [with transfer function $G_l(s)$] - I will call this system the laser (for my purpose), though for now it is modeled by the low pass filter. 

The controller essentially receives a signal corresponding to the deviation of the output from the input, and with this signal, the controller generates an input to the laser to drive the output towards the input. To achieve good performance, the controller must be optimized for the given system. By analyzing the block diagram and writing the loop equation, we see that $e(s) = R(s) − C(s)$ and $C(s) = G_l(s)G_c(s)e(s)$. The transfer function of the entire system from $R(s)$ to $C(s)$ is:
\begin{equation}
    T(s)\equiv \frac{\text{Output}}{\text{Input}}=\frac{C(s)}{R(s)}=\frac{G_c(s)G_l(s)}{1+G_c(s)G_l(s)}
\end{equation}
We can analyze this system and compare the response to that of the system without feedback (open loop).

\subsubsection{Transfer function of the laser}
\begin{figure}[hbtp]
\centering
\includegraphics[scale=0.25]{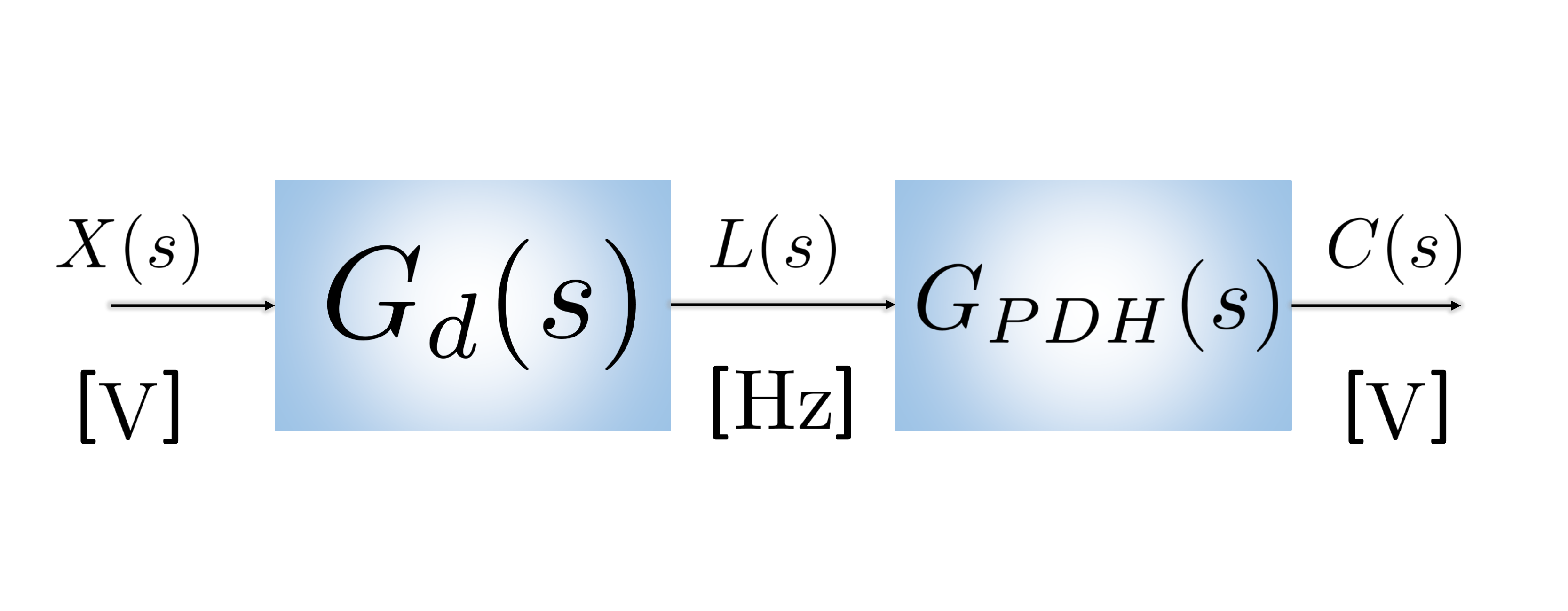}
\caption{The internal components of $G_l(s)$. The electrical signal $X(s)$ is converted to an optical frequency $L(s)$ in the diode which is determined using the PDH scheme and then converted back to an electric signal $C(s)$.}
\label{fig:glcomp}
\end{figure}

For the laser system, the signal is converted between different physical quantities as illustrated in figure \ref{fig:glcomp}. Let us imagine a diode laser where we control the output frequency, $L(s)$, by regulating the current in the diode, $X(s)$. We basically convert the electric signal to an optical frequency in $G_d(s)$, the transfer function of the diode. Whereas $X(s)$ was expressed in units of electric potential $(V)$, $L(s)$ is expressed in units of frequency (Hz). The conversion, $G_d(s)$, can itself have a non-trivial frequency dependence due the underlying semiconductor physics or limits in the electronics. The optical signal is then reflected on a
resonator and demodulated using the PDH technique. This process [with transfer function $G_{PDH}(s)$], converts the optical signal back to an electrical signal corresponding to $C(s)$ in figure \ref{fig:glcomp}. Though the internal structure is more complex than for the low pass filter, the system basically still have a an electrical input and output, and the system can be modeled in the same way. In principle, it is not even necessary to know the physical processes inside $G_l(s)$. For regulation purposes it suffices to know the transfer function itself, and this can be determined experimentally.

\begin{figure}[hbtp]
\centering
\includegraphics[scale=0.3]{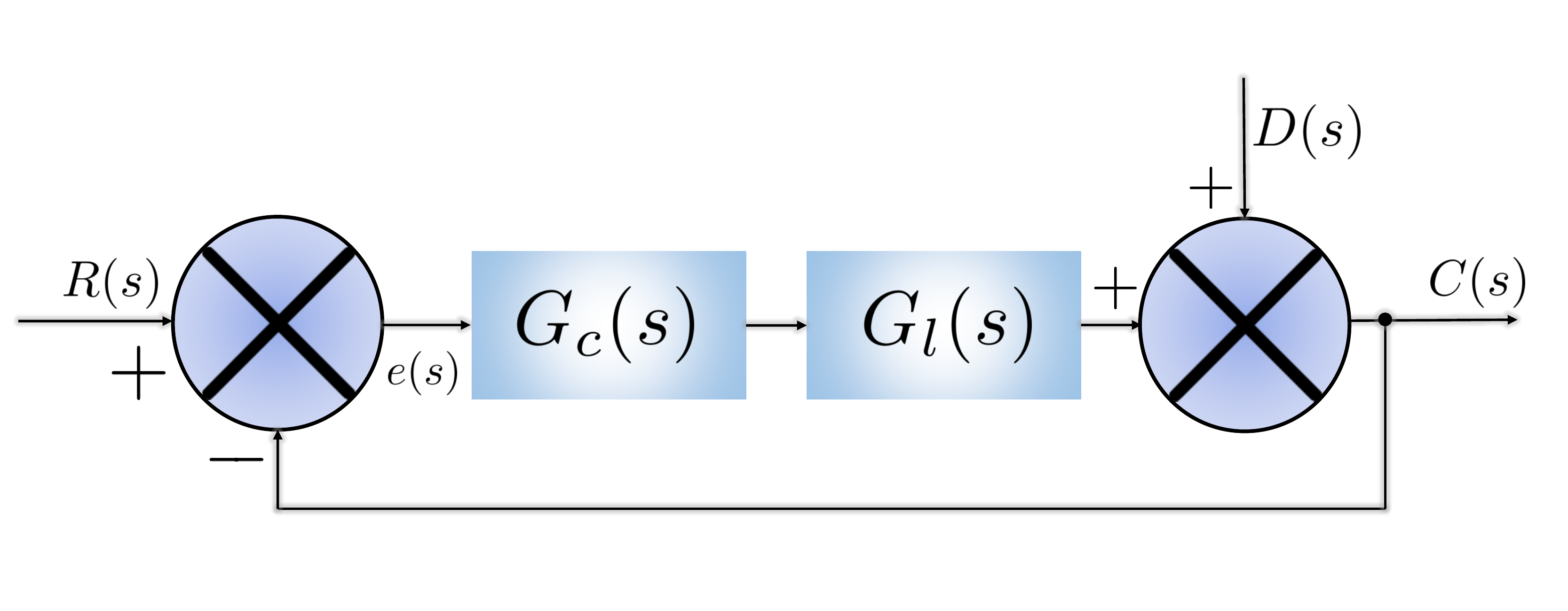}
\caption{Model of the control system for the laser locking. $D(s)$ models any disturbance in the laser frequency that drifts it away from resonance. It is the disturbance which the feedback system is supposed to suppress. Disturbance can be modelled in this way. In principle, it should be added to $L(s)$ as $C(s)$ is the actual PDH signal. For simplicity it can be effectively taken to be added this way. }
\label{fig:disturbance}
\end{figure}

For laser stabilization, the frequency is fixed and the feedback system should suppress unwanted disturbances which force the laser away from this frequency. The actual laser system can be modeled by introducing a disturbance, $D(s)$, to the system as illustrated in figure \ref{fig:disturbance}. The output of this system is:
\begin{equation}
    C(s)=\frac{R(s)G_c(s)G_l(s)+D(s)}{1+G_c(s)G_l(s)}
\end{equation}
When the disturbance is zero $[D(s) = 0]$, the transfer function of the system with respect to the input, $T(s) = C(s)/R(s)$. In laser stabilization, $R(s)$ is typically set to zero (i.e. the zero crossing of the PDH error signal) and we are interested in how a disturbance, $D(s)$, affects the system. In this case, it is the transfer function with respect to the disturbance which is relevant, given by:
\begin{equation}
    Q(s)=\frac{C(s)}{D(s)}=\frac{1}{1+G_c(s)G_l(s)}
\end{equation}
With the substitution $G(s) = G_c(s)G_l(s)$ and a slight rearrangement this can be re-written as
\begin{equation}
    Q(s)=1-\frac{G(s)}{1+G(s)}=1-T(s)
\end{equation}
The response of the system to a disturbance is very similar to the response to a change in input. The main difference is that the system finally settles to the new value when the input is changed, whereas a disturbance is eventually canceled out. This result shows that the standard techniques from control theory can be directly applied in the design of a system to cancel disturbances, such as in laser stabilization.

\subsubsection{PID controller as a feedback mechanism}
The obtained error signal needs to be treated (such as the scaling) to feed back to the controller. This can be achieved by the PID controller. The distinguishing feature of the PID controller is the ability to use the three control terms of proportional, integral and derivative influence on the controller output to apply accurate and optimal control. It continuously calculates an error value $e(t)$ as the difference between a desired setpoint SP$=r(t)$ and a measured process variable PV$=y(t)$, and applies a correction based on proportional, integral, and derivative terms. The controller attempts to minimize the error over time by adjustment of a control variable $u(t)$, such as the opening of a control valve, to a new value determined by a weighted sum of the control terms.

\begin{figure}[hbtp]
\centering
\includegraphics[scale=0.4]{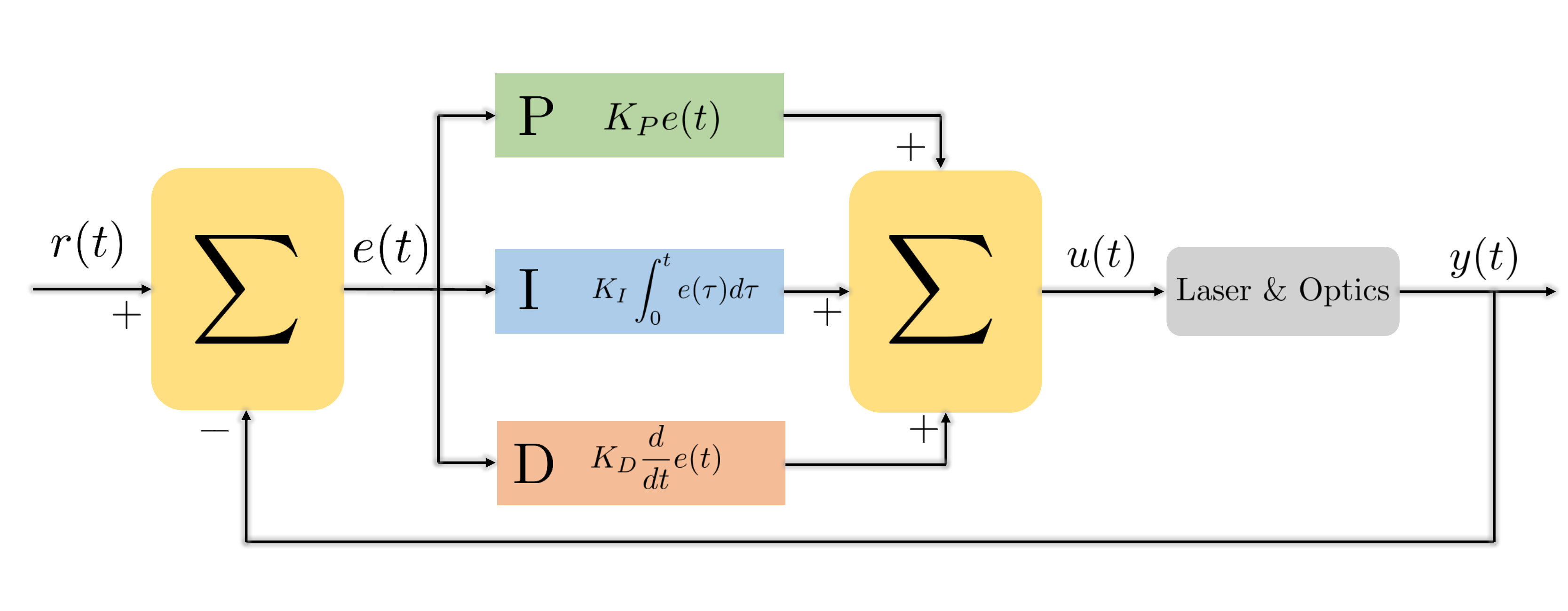}
\caption{A block diagram of a PID controller in a feedback loop. $r(t)$ is the desired process value or setpoint (SP), and $y(t)$ is the measured process value (PV).}
\label{fig:PID}
\end{figure}

Term \textbf{P} is proportional to the current value of the SP−PV error $e(t)$. For example, if the error is large and positive, the control output will be proportionately large and positive, taking into account the gain factor '$K$'. If there is no error, there is no corrective response. Term \textbf{I} accounts for past values of the SP−PV error and integrates them over time to produce the I term. For example, if there is a residual SP−PV error after the application of proportional control, the integral term seeks to eliminate the residual error by adding a control effect due to the historic cumulative value of the error. When the error is eliminated, the integral term will cease to grow. This will result in the proportional effect diminishing as the error decreases, but this is compensated for by the growing integral effect. Term \textbf{D} is a best estimate of the future trend of the SP−PV error, based on its current rate of change. It is effectively seeks to reduce the effect of the SP−PV error by exerting a control influence generated by the rate of error change. The more rapid the change, the greater the controlling or dampening effect.

The overall control function can be expressed mathematically as
\begin{equation}
    u(t)=K_Pe(t)+K_I\int_0^te(t')dt'+K_D\frac{d}{dt}e(t)
\end{equation}
where $K_P$, $K_I$, and $K_D$ (can be tuned to be positive or negative) denote the coefficients for the proportional, integral, and derivative terms respectively.

\subsubsection{PDH circuit and demonstration}
\begin{figure}[hbtp]
\centering
\includegraphics[scale=0.5]{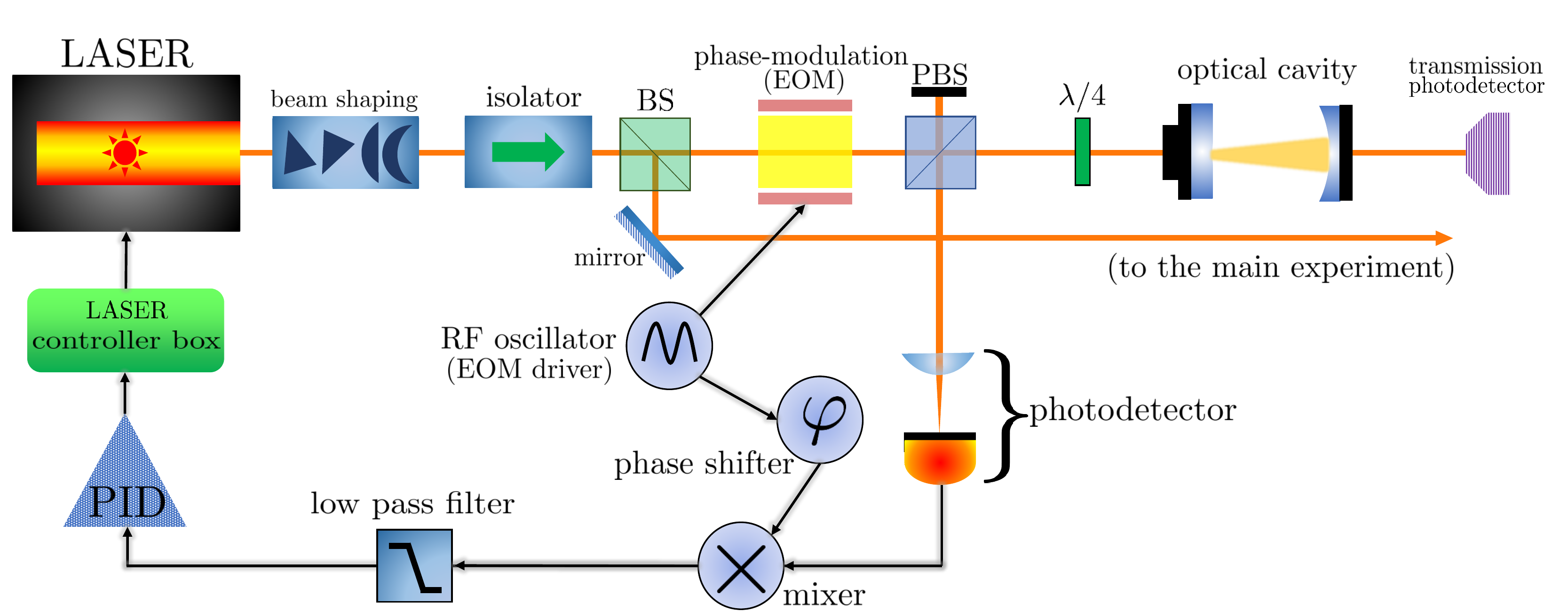}
\caption{Schematic diagram of the PDH technique used to stabilize a laser in a main experiment.}
\label{fig:PDHdiag}
\end{figure}

The follow figure, figure \ref{fig:PDHdiag} shows the whole setup of an experiment where PDH technique is used as a stabilization part of any main experiment (Ion trap in my case).
Starting point is the laser controller box or the laser driver which is used to control the laser frequency, the property we wish to stabilize.  In our experiment, we use an external-cavity diode laser (see Sec. \ref{subsec:diodelaser} for details on this type of laser). The laser beam coming out of the laser (which is usually linearly polarized in a certain plane) first passes through the beam shaping optical setup which usually consists of anamorphic prisms and some lens based telescopes to have a nice circular gaussian beam profile before the start of any experiment. The next is the isolator which prevents any back reflections from the optics to enter back into the laser (and might disrupt the laser mechanism and stability). These two components are not a part of the PDH setup but however, in practice, are extremely important. The beam then passes through a $50-50$ beam splitter (BS) where one part goes to the main experiment and other for the PDH stabilization. The PDH beam passes through the EOM (Electro-Optic-Modulator) which results in the phase modulation of the beam and in conclusion adds the sidebands to the beam. This EOM is driven by a RF-oscillator at a resonant frequency equal to the frequency $\Omega$ of the sidebands. The beam (which now has sidebands around the main carrier frequency) passes through a polarizing-BS (PBS) which sends a beam to the cavity and other part is irrelevant and is blocked. The straight beam passes through the $\lambda /4$ quarter waveplate and is incident on the cavity which reflects the sidebands and transmits the carrier when in resonance. The transmitted signal (the FSR peaks) fall on the transmission detector which can be used to characterize the cavity. The reflected beam off the cavity passes again through the quarter waveplate again (effectively a half-waveplate with a $\pi/2$ turned direction of linear polarization to effectively change the direction of propagation) and reflects off in the other direction on the photodetector which captures the error signal and converts it to a electronic signal. This error signal is fed to the mixer along with the RF signal which comes from the same RF-oscillator which drives the EOM through a phase shifter with variably controlled phase delay. This signal passes through a low pass filter to extract the error signal. This error signal is sent to the PID controller to convert to the right feeedback signal for the controller to drive the laser drift when the laser misbehaves off the resonance. The controller box takes the feedback from the PID controller and changes the current, temperature and the piezo to stabilize the laser.

It is not mandatory that the right laser frequency needed exactly lies on the FSR points of the cavity. If this is not the case (which most often is) then we require to bridge the gap from the nearest FSR point to the laser frequency to obtain the right stabilized laser frequency for our experiment (in my case this is the 866 nm or 346 THz and the nearest FSR point was just around 500 MHz away from this). This is achieved by the use of AOM (Acoustic-Optical-Modulator) to bridge the gap to the right frequency needed.

The PDH technique benefits from the fact that the error signal on resonance is independent of the optical power. This is important to avoid offset changes which could otherwise change the frequency of the laser.
Fluctuations in the optical power will, however, change the slope of the error signal and hence the gain in the feedback loop. This can affect the performance of the lock, and power fluctuations should therefore be kept to a minimum.
 
The PDH technique in itself is insensitive to changes in the polarization, but if polarization dependent optical components are used, these can result in a modified error signal. It is particularly worth noting that if the polarization of the light is not perfectly matched to the axes of the EOM crystal, residual amplitude modulation can appear after polarization dependent elements. This modulation will oscillate with frequency $\Omega$ and therefore give rise to an offset in the lock signal. For a constant offset (constants amplitude modulation) this can simply be compensated electronically. With a fluctuating offset this will however translate into fluctuations in the frequency of the light. It is therefore advantageous to obtain a clean polarization before the EOM; for example with a \textit{Glan-Thompson} Polarizer.

The modulation frequency, $\Omega$, can in principle be chosen arbitrarily but it will limit the operating range and bandwidth of the final system. Instantaneous frequency fluctuations larger than $\Omega$ will give rise to an error signal with the wrong sign and cannot be corrected - the laser jumps out of lock. Additionally, the low pass filter used to isolate the amplitudes in front for the harmonic $\Omega$ terms will limit the bandwidth of the system. To correct high frequency fluctuations it is required to use a high modulation frequency.

\subsection{Fundamental limits and noise}
\subsubsection{Parameter noise}
The main contribution to noise comes from the variation of the cavity length. Any acoustic or thermal changes can bring a minute change in the cavity length and change the error signal that is obtained. Ideally, the cavity must be stored in a temperature controlled box in isolation. Change in the frequency or the length of the cavity are both on equal footings for the PDH technique near the resonance. For the high modulation frequencies
\begin{equation}
    \epsilon=-8\sqrt{P_cP_s}\frac{2L\mathscr{F}}{\lambda}\left[\frac{\delta \omega}{\omega}+\frac{\delta L}{L}\right]
\end{equation}
where $\delta L$ is the drift of the length of the cavity from the resonance similar to $\delta \omega$. This can conversely understood as stabilizing the cavity using a laser (assuming that the laser is a well stabilized already by some other locking), just by switching roles of the length and the frequency \cite{ref:14}. Thus, it is not possible to distinguish between the cavity noise and the laser frequency noise in PDH by the analysis of the error signal alone. 

There are few other factors to be taken into account that affect or contribute to the error signal. None of the following contribute to the error signal to \textit{first order}: variation in the laser power, response of the photodiode used to measure the reflected signal, the modulation depth $\beta$, the relative phase of the two signals going into the mixer, and the modulation frequency $\Omega$. The system is insensitive to each of these because we are locking on resonance, where the reflected carrier vanishes. This causes all of these first-order terms to vanish in a Taylor expansion of the error signal about resonance. The system is first-order sensitive to fluctuations in the sideband power at the modulation frequency $\Omega$. Most noise sources fall off as frequency increases, so we can usually reduce them as much as we want by going to a high enough modulation frequency. There is one noise source, however, that does not trail off at high frequencies, and that is the shot  noise in the reflected sidebands. Shot noise has a flat spectrum, and for high enough modulation frequencies it is the 
dominant noise source.

\subsubsection{Shot noise and fundamental limits for resolution (Quantum noise limit)}
Any noise in the error signal itself is indistinguishable from noise in the laser’s frequency. There is a fundamental limit to how quiet the error signal can be, due to the quantum nature of light. 
Shot noise describes the fluctuations of the number of photons detected (or simply counted in the abstract) due to their occurrence independent of each other. This is therefore another consequence of discretization, in this case of the energy in the electromagnetic field in terms of photons. In the case of photon detection, the relevant process is the random conversion of photons into photo-electrons for instance, thus leading to a larger effective shot noise level when using a detector with a quantum efficiency below unity. The shot noise of a coherent optical beam (having no other noise sources) is a fundamental physical phenomenon, reflecting quantum fluctuations in the electromagnetic field. 

On resonance, the reflected carrier will vanish, and only the sidebands will reflect off the cavity and fall on the photodetector. These sidebands will produce a signal that oscillates at harmonics of the modulation frequency. Calculating the shot noise in such a cyclostationary signal is fairly subtle, but for the purpose it can be estimated by replacing this cyclostationary signal with an averaged, dc signal. The average power falling on the photodiode is $P_{ref}=2P_s$ approximately. The shot noise in this signal has a flat spectrum with spectral density of
\begin{equation}
    S_e=\sqrt{2\frac{hc}{\lambda}(2P_s)}
\end{equation}
Dividing the error signal spectrum by $D$ (frequency determinant) gives us the apparent frequency noise \cite{ref:14},
\begin{equation}
    S_f=\frac{\sqrt{hc^3}}{8}\frac{1}{\mathscr{F}L\sqrt{\lambda P_c}}
\end{equation}
Since we can’t resolve the frequency any better than this, we can’t get it any more stable than this by using feedback to control the laser. Note that the shot noise limit does not explicitly depend on the power in the sidebands. It only depends on the power in the carrier. The same shot noise would limit the sensitivity to cavity length if we were locking the cavity to the laser.

\newpage
\section{Experimental Realization}
In this section, I describe my implementation of the PDH technique. I describe the equipment used and their brief working along with the related results obtained.
The following figure, figure \ref{fig:OpticalCircuit} shows the full implementation of the setup. I have built the PDH part on a separate optic plate which can be plugged into any main experiment (for 866nm). All the optics used in the experiment are from \href{https://www.thorlabs.com/}{ThorLabs}. All the labelled components are shown in the figure itself. This is the actual experimental realization of the schematic circuit shown in figure \ref{fig:PDHdiag}. We start with a laser and follow with the beam shaping, isolator and finally a PBS and a l/2 waveplate (to control the power) which gives a variable power in two parts, one for the wavlength readout (which typically takes only about 5$\%$ of the total power and about 10 mW is sent to the PDH circuit). Here, main experiment is not shown. But a BS can be put anywhere before the previous PBS to direct one part to the main experiment. To enter the cavity, a fiber is used rather than direct path using mirrors. This is done to gain the freedom for the cavity PDH circuit. It is built on a separate optical slab (the black one in the diagram) so that it can moved independently to any place in the lab without needing to reconstruct everything. This fiber is the polarization maintaining fiber and must be entered with a correct polarization (otherwise it results in a circularly polarized light at the emission end). Two waveplates - l/2 and l/4 are also used before the EOM. This is done to enter the EOM again with the right polarization. The configuration of these waveplates can be best chosen by analyzing the output of the beam emerging after the EOM. This EOM results in two beam spots (see figure \ref{fig:EOM}, separated by angular distance of about 4$^\circ$), from which the highly deviating (the lower one in this case) must be blocked with the help of an iris (placed just before the cavity). Then a BS is used (without a l/4 waveplate) for entering the cavity. Ideally, a PBS and a l/4 waveplate must be used. But the use of isolator prevents from any back scattering into the laser. One part reflects off and is blocked using a blocker. Other part is incident on the  optical cavity. The resonant frequency transmits to the camera and the photodetecor. Transmitted signal is viewed in the computer using the beam-profiler and and with the oscilloscope using the signal from the photodetector, this gives complete transmission information. 

\begin{figure}[hbtp]
\centering
\includegraphics[scale=0.5]{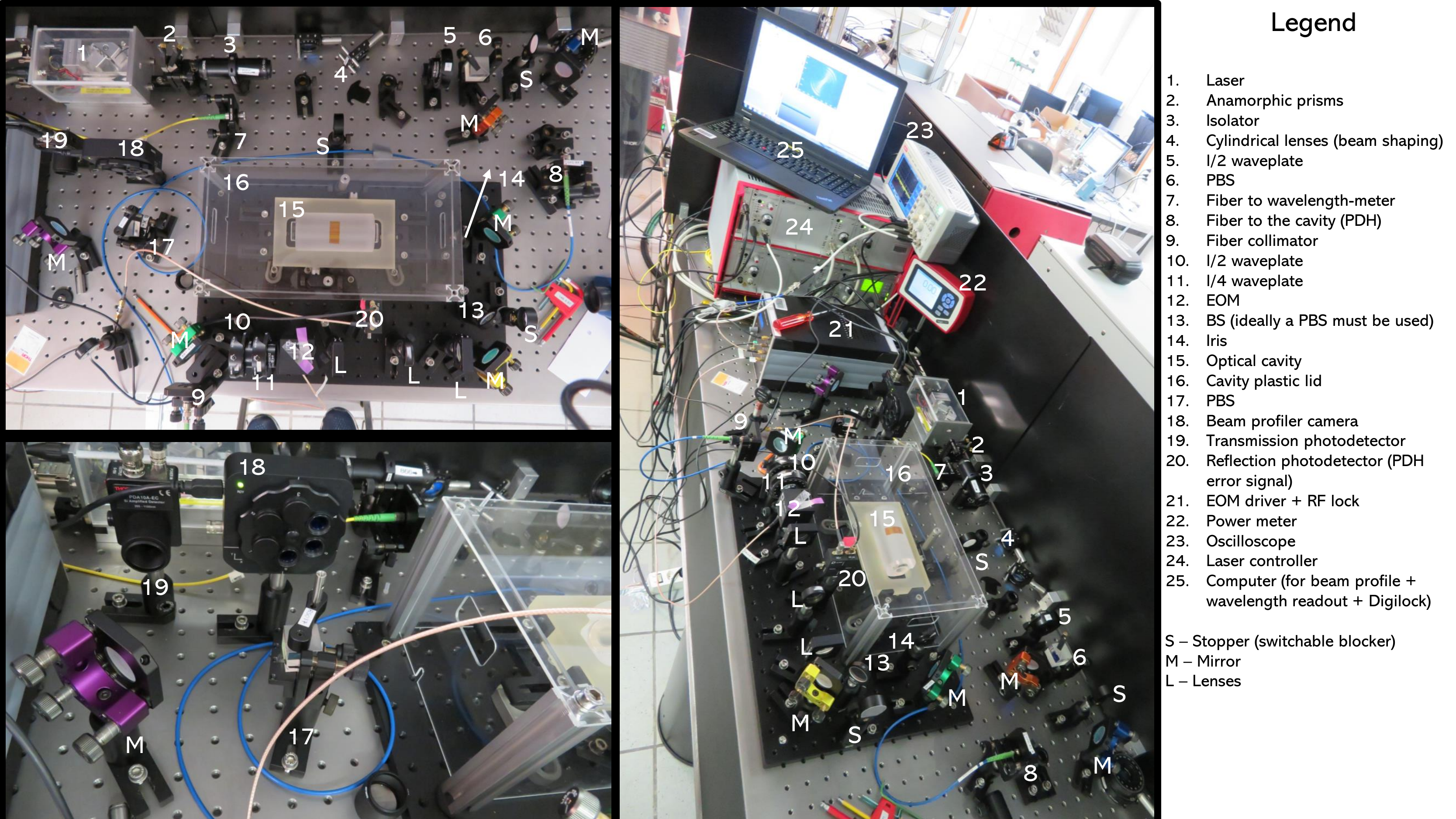}
\caption{The experimentally constructed PDH optical setup.}
\label{fig:OpticalCircuit}
\end{figure}
The sidebands reflect off the cavity and again reflect off the BS and are incident on the reflection photodetector which sends the PDH reflection signal to the RF lock box which outputs a final PDH error signal which is input to the PID(s). The PID output after the necessary treatment goes into the laser controller box for modulation and feedback. The rest of the circuit is exactly same as explained in the theoretical description.

\subsection{Diode laser (Grating stabilized 866nm)}
\label{subsec:diodelaser}
\begin{figure}[hbtp]
\centering
\includegraphics[scale=0.4]{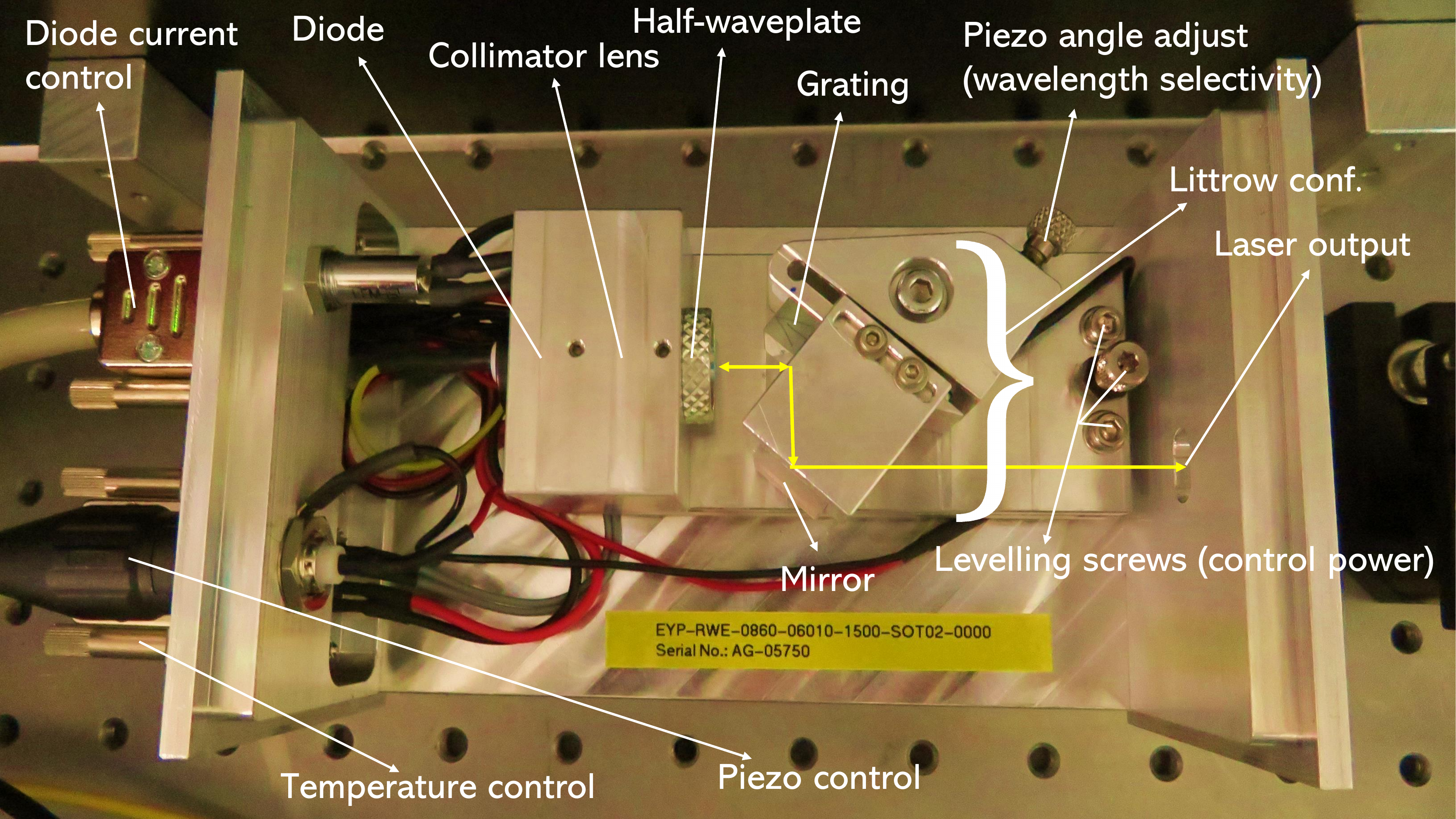}
\caption{Schematic of the self-built grating stabilized 866nm diode laser with Littrow configuration. The mirror is only used to reflect the beam from the grating to change direction.}
\label{fig:diodelaser}
\end{figure}

The active region of the laser diode is in the intrinsic (I) region, and the carriers (electrons and holes) are pumped into that region from the N and P regions respectively. When an electron and a hole are present in the same region, they may recombine or "annihilate" producing a spontaneous emission — i.e., the electron may re-occupy the energy state of the hole, emitting a photon with energy equal to the difference between the electron's original state and hole's state. Spontaneous emission below the lasing threshold produces similar properties to an LED. Spontaneous emission is necessary to initiate laser oscillation, but it is one among several sources of inefficiency once the laser is oscillating. Due to diffraction, the beam diverges (expands) rapidly after leaving the chip. For my experiment, an external cavity is used to achieve 'lasing' and also select the wavelength to 866 nm by the use of a piezo. This is a self-built laser in the lab and is shown in figure \ref{fig:diodelaser}.
\begin{figure}[hbtp]
\centering
\includegraphics[scale=0.5]{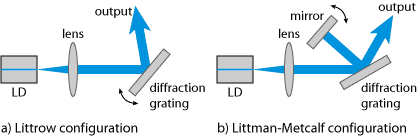}
\caption{Tunable external-cavity diode lasers in Littrow and Littman–Metcalf configuration. \textit{Image credits: \href{https://www.rp-photonics.com/external_cavity_diode_lasers.html}{RP photonics}}}
\label{fig:conf}
\end{figure}

External-cavity diode lasers (ECDL) are tunable lasers which use mainly double heterostructures diodes. The first external-cavity diode lasers used intracavity etalons and simple tuning Littrow gratings (figure \ref{fig:conf}). Other designs include gratings in grazing-incidence configuration and multiple-prism grating configurations. The first-order diffracted beam provides optical feedback to the laser diode chip, which has an anti-reflection coating. The emission wavelength can be tuned by rotating the diffraction grating. In the Littman–Metcalf configuration, the grating orientation is fixed, and an additional mirror is used to reflect the first-order beam back to the laser diode. The wavelength can be tuned by rotating that mirror. This configuration offers a fixed direction of the output beam, and also tends to exhibit a \textit{smaller} linewidth, as the wavelength selectivity is stronger. (The wavelength-dependent diffraction occurs twice instead of once per resonator round trip.) A disadvantage is that the zero-order reflection of the beam reflected by the tuning mirror is lost, so that the output power is lower than that for a Littrow laser where no mirror is used and the reflection occurs only once. The angle of the diffraction grating is fine tuned by the use of piezoelectric crystal which is driven by the controller. Upon the application of voltage, the crystal expands and contracts which changes the angle of the grating. Because very high voltages correspond to only tiny changes in the width of the crystal, this crystal width can be manipulated with better-than-micrometer precision. The laser I used has a Littrow configuration.

\begin{figure}[hbtp]
\centering
\includegraphics[scale=0.23]{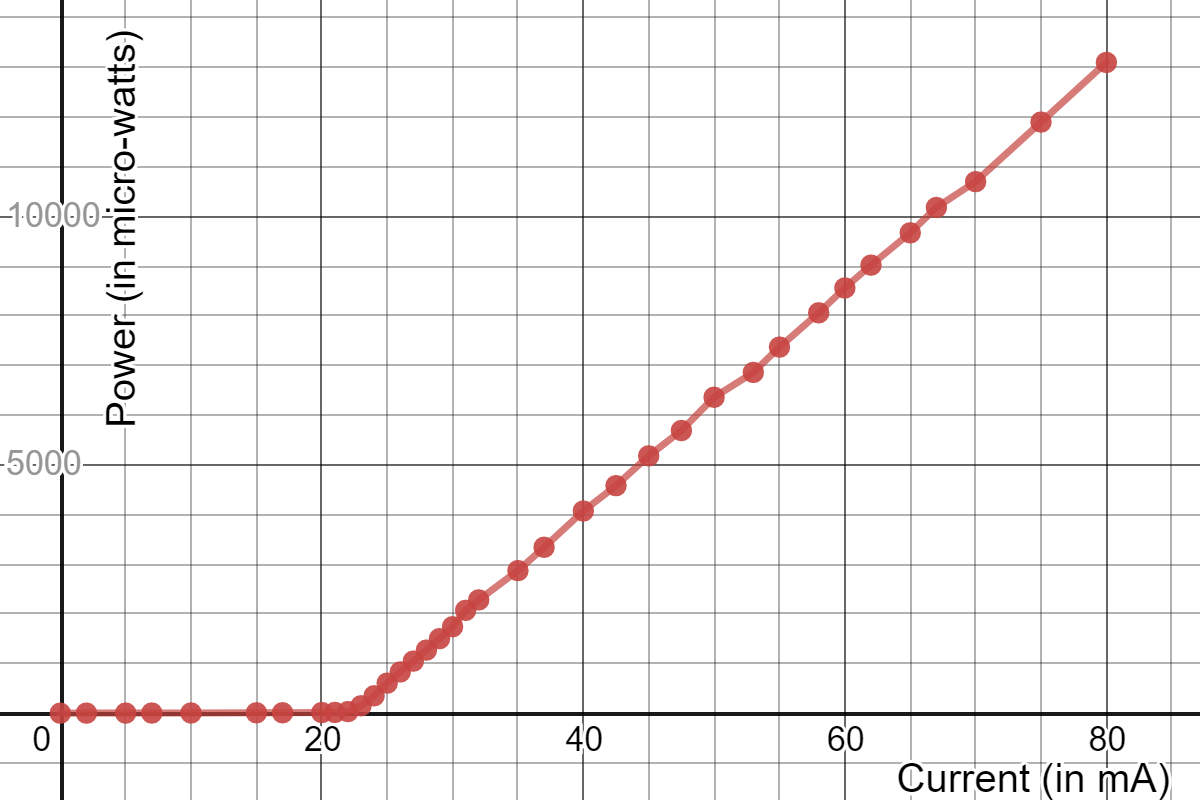}
\caption{PI curve of the laser at $20.4^\circ$ C.}
\label{fig:PIcurve}
\end{figure}

\begin{figure}[hbtp]
\centering
\begin{subfigure}
\centering
\includegraphics[scale=.80]{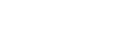}
\end{subfigure}
\begin{subfigure}
\centering
\includegraphics[scale=0.6]{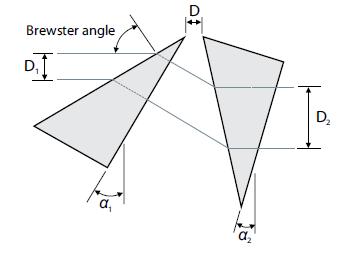}
\end{subfigure}
\begin{subfigure}
\centering
\includegraphics[scale=1.0]{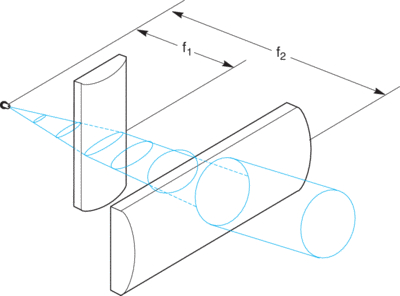}
\end{subfigure}
\caption{Beam shaping with the help of anamorphic prisms and cylindrical lenses. Any elliptical beam profile can be made fine and circular with these arrangements.}
\label{fig:beamshaping}
\end{figure}
The diode the laser is built for is \href{https://www.eagleyard.com/fileadmin/downloads/data_sheets/EYP-RWE-0860-06010-1500-SOT02-0000.pdf}{EYP-RWE-0860-06010-1500-SOT02-0000} which is driven at around 64mA of current and $20.4^\circ$ C and gives an output power of 10 mW at this current with the current setup. The beam profile of the diode is elliptical and nearly horizontal polarized. The PI curve of the laser is shown in figure \ref{fig:PIcurve}. The laser starts lasing at around 22 mA. Next comes the beam shaping. The beam profile is made circular with the help of the anamorphic prisms and a cylindrical lens telescope (optional, just an improvement. In principle, only anamorphic prisms are enough, figure \ref{fig:beamshaping}). 

The principle of beam shaping with anamorphic prism pairs is not based on focusing effects (i.e., changes of wavefront curvature), but rather on changes of the beam radius for refraction at flat prism interfaces. Such changes occur at the interfaces of any prism (except for normal incidence), because the angle of the beam against the surface-normal direction is different inside and outside according to Snell's law. However, for a symmetric beam path, where the beam angles against input and output face of the prism are identical, the two changes in beam radius cancel each other. Therefore, one has to use an asymmetric configuration. A single prism is sufficient for changing the beam radius in one direction, but it also changes the beam direction. By using an anamorphic prism pair, one can obtain an output beam with an unchanged direction, only a position offset. The two prisms are of course oriented such that they change the beam radius in the same direction. The overall magnification is then the square of the refractive index, or the inverse of that.

\subsection{Controller box}
\label{subsec:controller}
\begin{figure}[hbtp]
\centering
\includegraphics[scale=0.45]{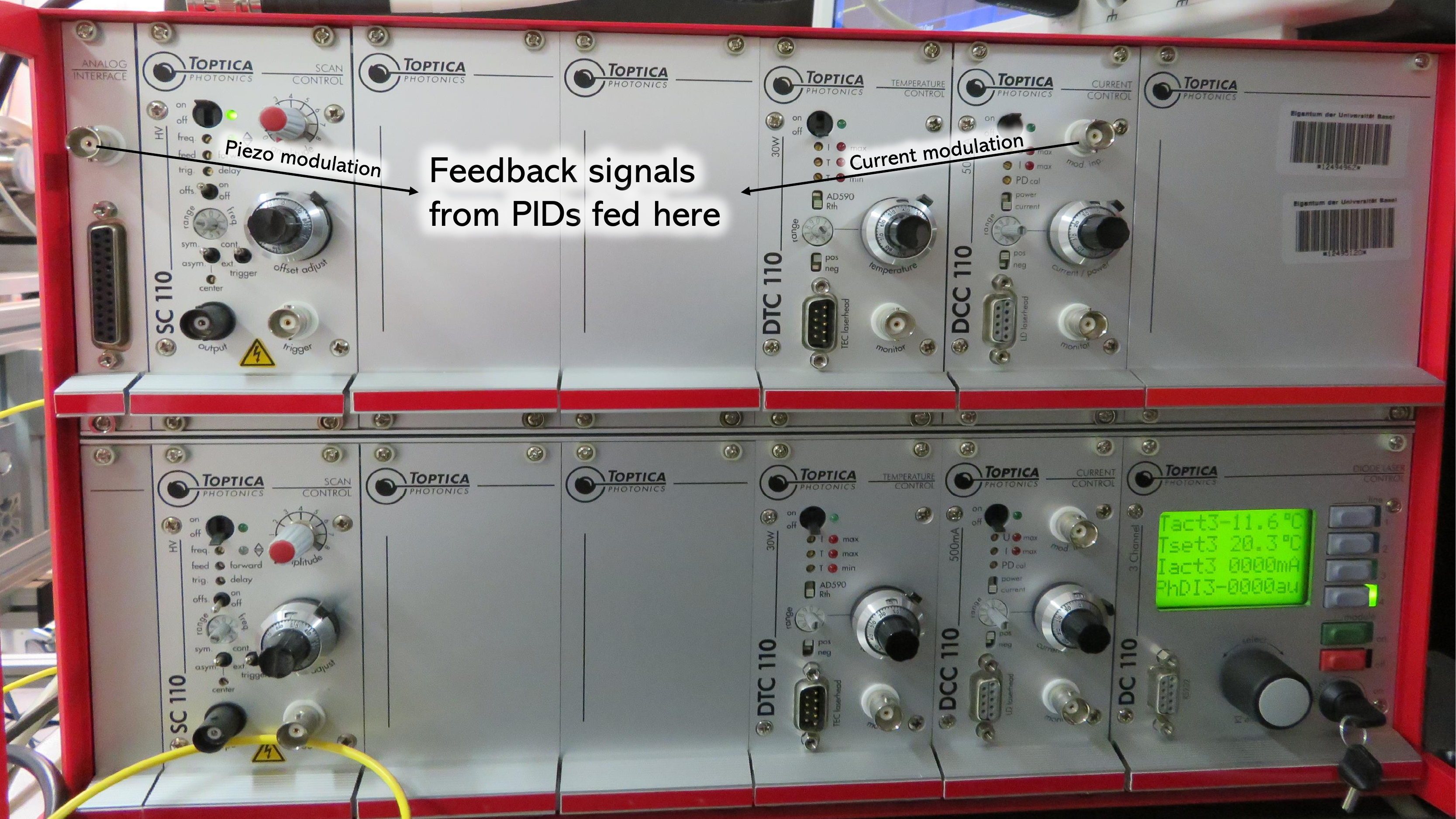}
\caption{The front panel and interface of the controller box.}
\label{fig:controller}
\end{figure}

The controller (laser driver) that I have used in my experiment is \href{https://www.toptica.com/products/tunable-diode-lasers/laser-driving-electronics/sys-dc-110-analog-control/}{SYS-DC 110 Analog-Control by Toptica (500 mA version)}. The front interface is shown in figure \ref{fig:controller}. It has a separate current driver and temperature control. The scan control is used for the piezo driving and enables the fine selection of wavelength/frequency of the order of 1MHz along with variable speed scan of frequency. The controller also enables the input for a feedback (from a PID, for example, which I have used) to stabilize the piezo, temperature and current independently.

There are two modes for the piezo control, the high voltage (HV) and the low voltage (LV) mode. The output piezo voltage in HV is -5 V to +150 V and in LV mode it is -12 V to +12 V (this corresponds to about 5 FSR equivalent $\approx$ 7 GHz over the whole range in LV mode and gives the least count to be about 2.4 mV). These modes can be selected by the jumpers inside the piezo panel. For the feedback at the end of the PDH technique, two separate feebacks from two PIDs can be fed to the piezo and the current control separately via BNC connectors. The bandwidth of the controller allows an output modulation. The sensitivity is +1 V for +10 mA change for current modulation in LV mode. 

\subsection{Fiber coupling}
There are two types of optic fibers, the single mode and the multimode. Multi-mode fiber has a fairly large core diameter that enables multiple light modes to be propagated and limits the maximum length of a transmission link because of modal dispersion. The main difference between multi-mode and single-mode optical fiber is that the former has much larger core diameter, typically 50–100 micrometers; much larger than the wavelength of the light carried in it. Because of the large core and also the possibility of large numerical aperture, multi-mode fiber has higher light-gathering capacity than single-mode fiber. In practical terms, the larger core size simplifies connections. In contrast, the lasers used to drive single-mode fibers produce coherent light of a single wavelength. Because of the modal dispersion, multi-mode fiber has higher pulse spreading rates than single mode fiber, limiting multi-mode fiber’s information transmission capacity. Single-mode fibers are often used in high-precision scientific research because restricting the light to only one propagation mode allows it to be focused to an intense, diffraction-limited spot. Some advantages of the single-mode fiber are no degradation of signal and low dispersion but on the expense that it is harder to couple light into the fibre and needs very precise optomechanical instruments. In order to couple light of wavelength $\lambda$ from a collimated laser beam of $1/e^2$ diameter $D$ into a fiber of mode field diameter $\eta$, choose a lens with a focal length $f = D( \pi \eta/4\lambda)$. There are also special kinds of fibers such as the Polarization-Maintaining (PM) fibers. I used a PM fiber to send the light to the cavity, since the EOM before the cavity requires a particular axis entry of linearly polarized light. Optical fibers are a flexible tool that allow to separate the experiment from the laser setup. Nevertheless it has been observed that, due to temperature and pressure fluctuation or just mechanical stress on the cable, the laser linewidth can be be changed up to some kHz. This phase noise is more evident for longer fibers.
\begin{figure}[hbtp]
\centering
\includegraphics[scale=0.4]{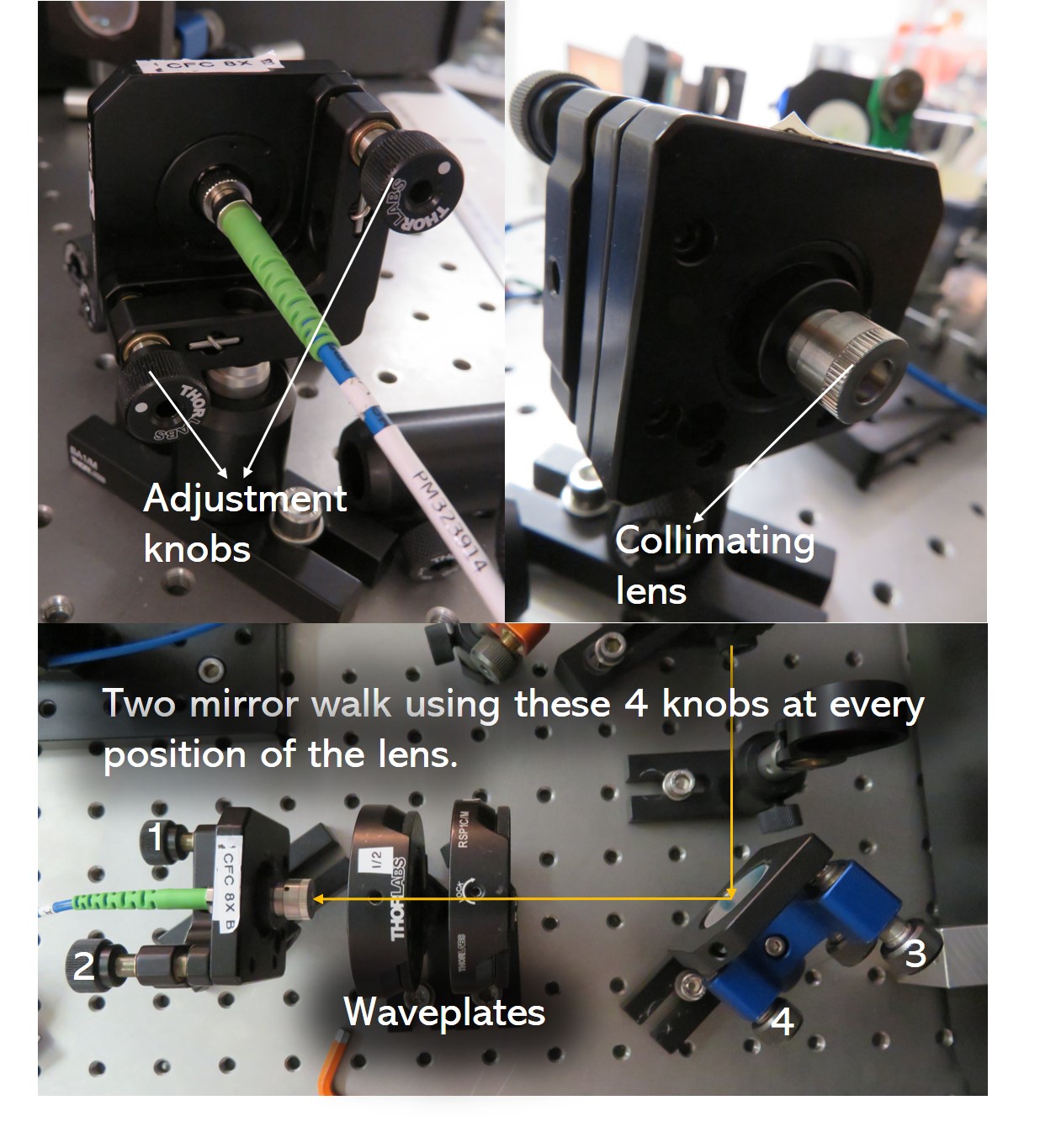}
\caption{Rear and front view of the two-axis fiber coupler with the collimating lens along with the depiction of the two mirror walk method.}
\end{figure}

The best coupling of the fiber in my experiment was about 55$\%$ power ratio. In brief, first I set a rough optical alignment manually. Then I send a visible laser light into the coupler in opposite direction and match its axis with the incoming infrared light (866nm) by the 'two mirror walk' (actually a coupler and a mirror here). And then insert the fiber into the coupler and perform again the two mirror walk with the other end of the fibre connected to a power-meter to maximize the coupling. Usually a collimating lens is needed for a particular beam cross-section size. So, with every coordinate of the collimating lens, one has to perform the two mirror walk in order to gain the best coupling.

\subsection{EOM and the sidebands}
\label{subsec:EOM}
\begin{figure}[hbtp]
\centering
\includegraphics[scale=0.45]{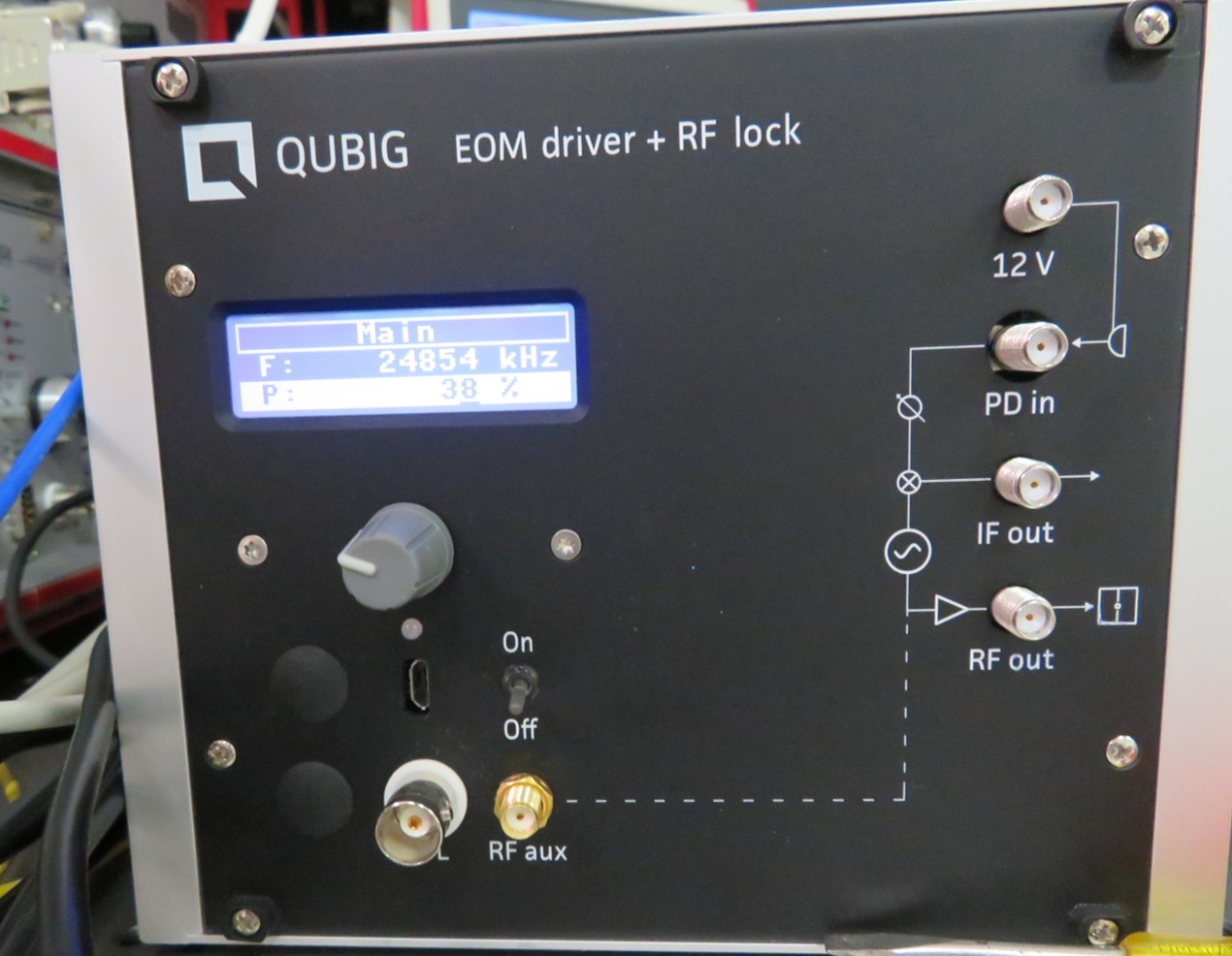}
\caption{EOM driver and the RF lock. It has inbuilt mixer, phase shifter and the low pass filter to extract the final PDH signal from the IF-out port which is fed to the PID.}
\label{fig:RFDriver}
\end{figure}

\begin{figure}[hbtp]
\centering
\includegraphics[scale=0.35]{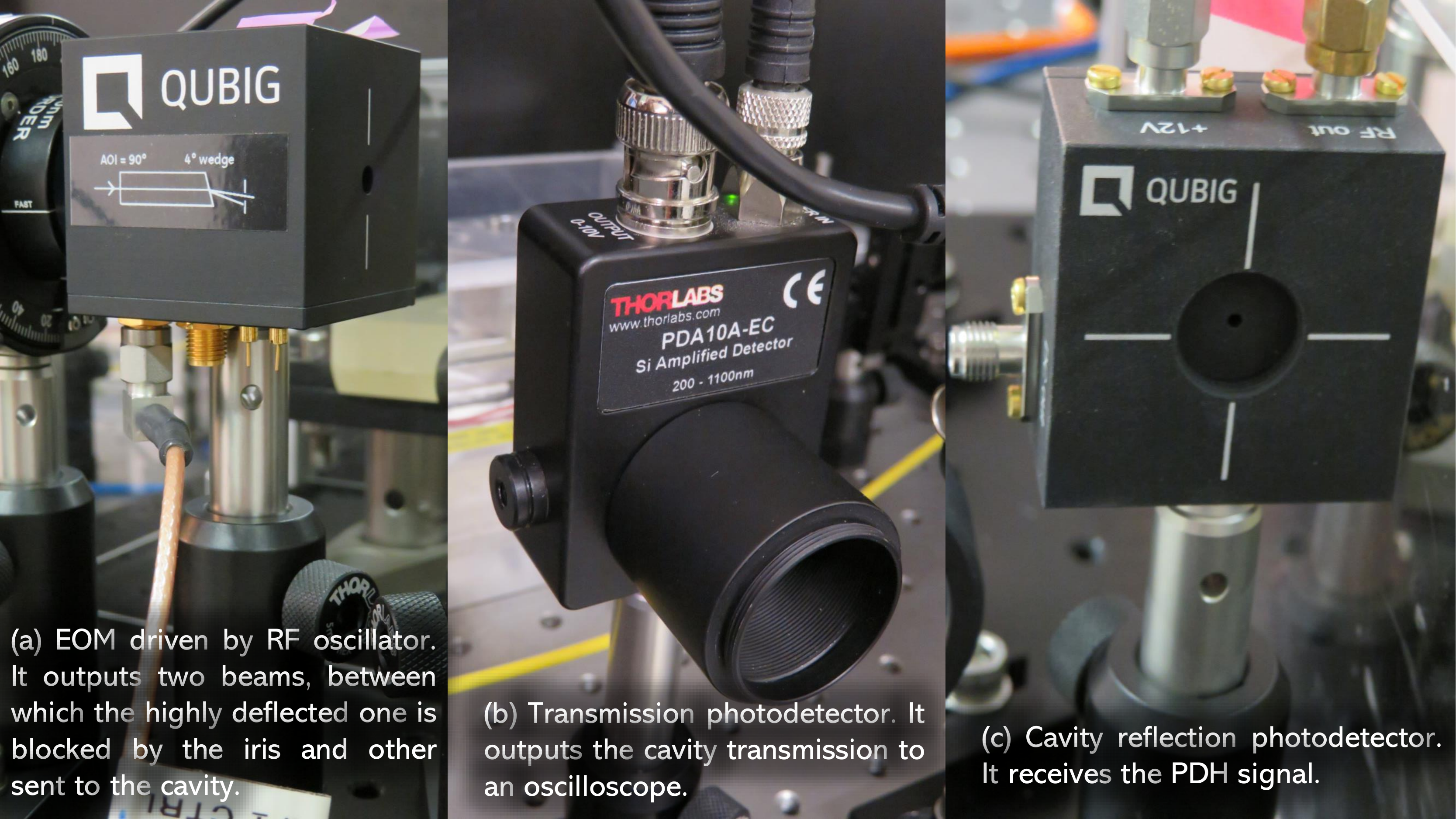}
\caption{EOM crystal at RF frequency, transmission detector and the PDH signal photodetector.}
\label{fig:EOM}
\end{figure}

\begin{figure}[hbtp]
\centering
\includegraphics[scale=0.48]{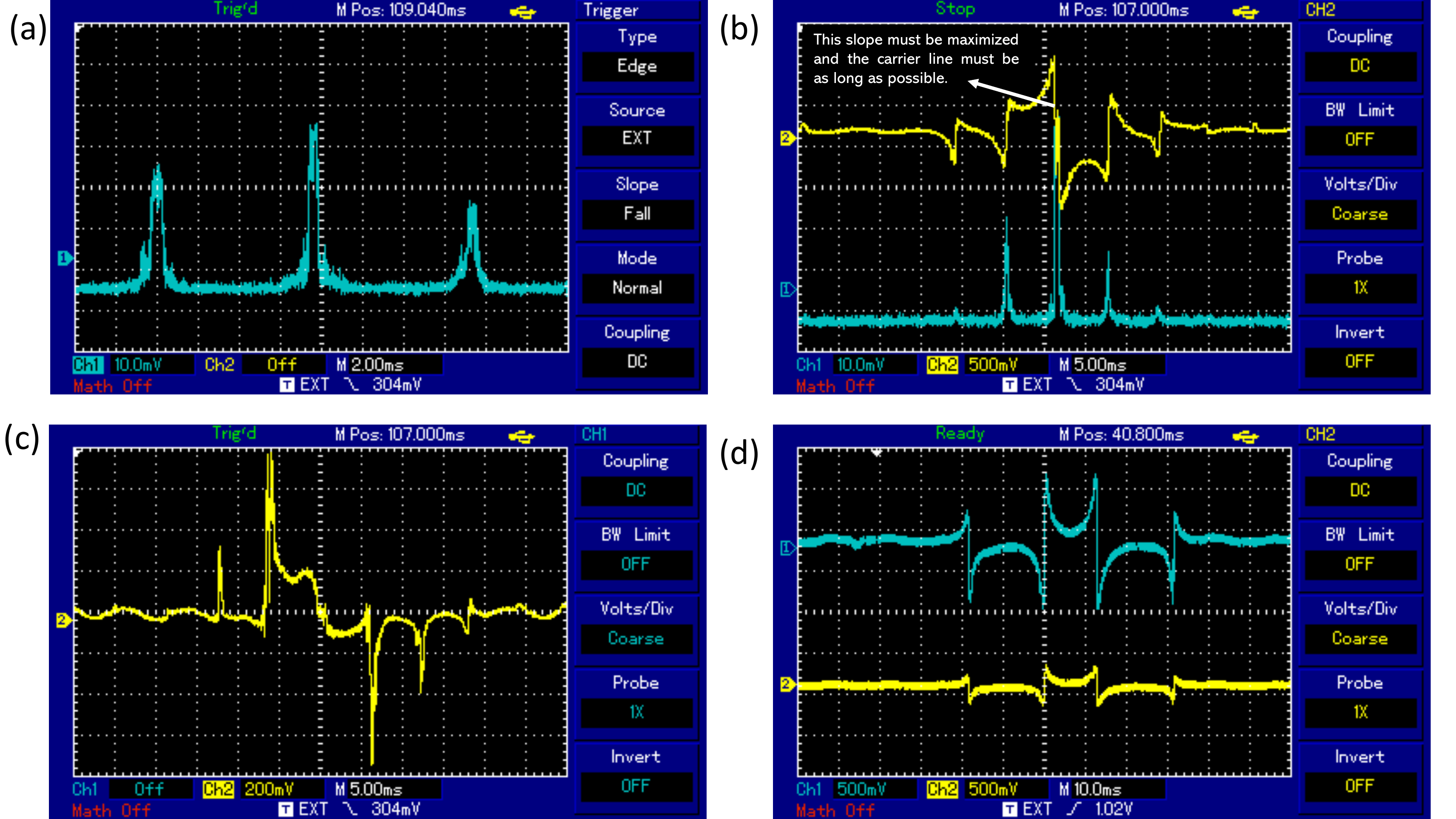}
\caption{Various obtained signals for transmission and reflection: (a) The sidebands obtained when the EOM is driven with $\Omega$ and the frequency is scanned. Two sidebands appear at equal spacing. The sidebands can be modulated by their power. (b) The PDH signal (yellow) with the sidebands (blue), the slope around the carrier is the key to the whole technique here. This must be optimized using the phase and the power to have the maximum slope and the curve to have the maximum length of linear form near the resonance. (c) PDH signal waveform changes completely when the phase is changed using the phase shifter inside the RF lock box. In this example, this phase is not suitable for the PDH technique. A PDH signal like in (b) can be regarded as a good PDH signal. (d) The output of the PID (yellow) and the original PDH signal (blue). The PID parameters can be changed in order to have the best feedback signal. Here only \textbf{P} and \textbf{I} are shown.}
\label{fig:PDHScope}
\end{figure}

For the phase modulation, transmission detection and the PDH signal photodetection I used the \href{https://www.qubig.com/products/electronics/rf-driver.html}{QUBIG EOM driver + RF lock} $\&$ \href{https://www.qubig.com/products/electro-optic-modulators-230/phase-modulators/pm7-nir.html}{QUBIG EOM phase modulator PM7-NIR} and the \href{https://www.qubig.com/products/electronics/photodiodes/pd-nir.243.html}{QUBIG PD-NIR photodetector}. They are shown in figure \ref{fig:RFDriver} and figure \ref{fig:EOM} respectively. The EOM driver box has an RF oscillator driver which drives the EOM at around 28 MHz (for my experiment) and also an inbuilt phase shifter, mixer and a low pass filter. The output of the PDH photodetector is fed to the input of the mixer along with the other inbuilt RF input which gives the error signal which can be treated with the phase shifter built in. The output of this PDH-lock box goes to the oscilloscope (to modify the signal to best need in real-time) and the PID controller which treats the signal for the feedback. The sidebands generated by the EOM can be visualized by scanning the frequency and triggering the transmission photodetector signal on the oscilloscope. It is shown in figure \ref{fig:PDHScope}(a). The power of the sidebands can be controlled using the RF driver box. When the proportional power is increased, the sidebands grow in size consuming the power from the carrier peak and vice versa. After the EOM and some optics, the laser beam enters the cavity.

At first, the EOM should only be aligned and not powered. The cavity has to be aligned first and its characteristics must be estimated (for a better scientific approach). The transmission via the cavity must be maximized and the TEM$_{00}$ mode must be found without a mode-hop in the neighbourhood frequency domain. And resonance must be made as much close as possible to the required wavelength of 866.451700 nm (vacuum wavelength for the D$_{3/2}$ ↔ P$_{1/2}$ transition resonance in $^{40}$Ca$^+$). After this is achieved, the EOM drive must be switched on and the sidebands must be analyzed.

\subsection{Optical cavity specifications and alignment}
\begin{figure}[hbtp]
\centering
\includegraphics[scale=0.34]{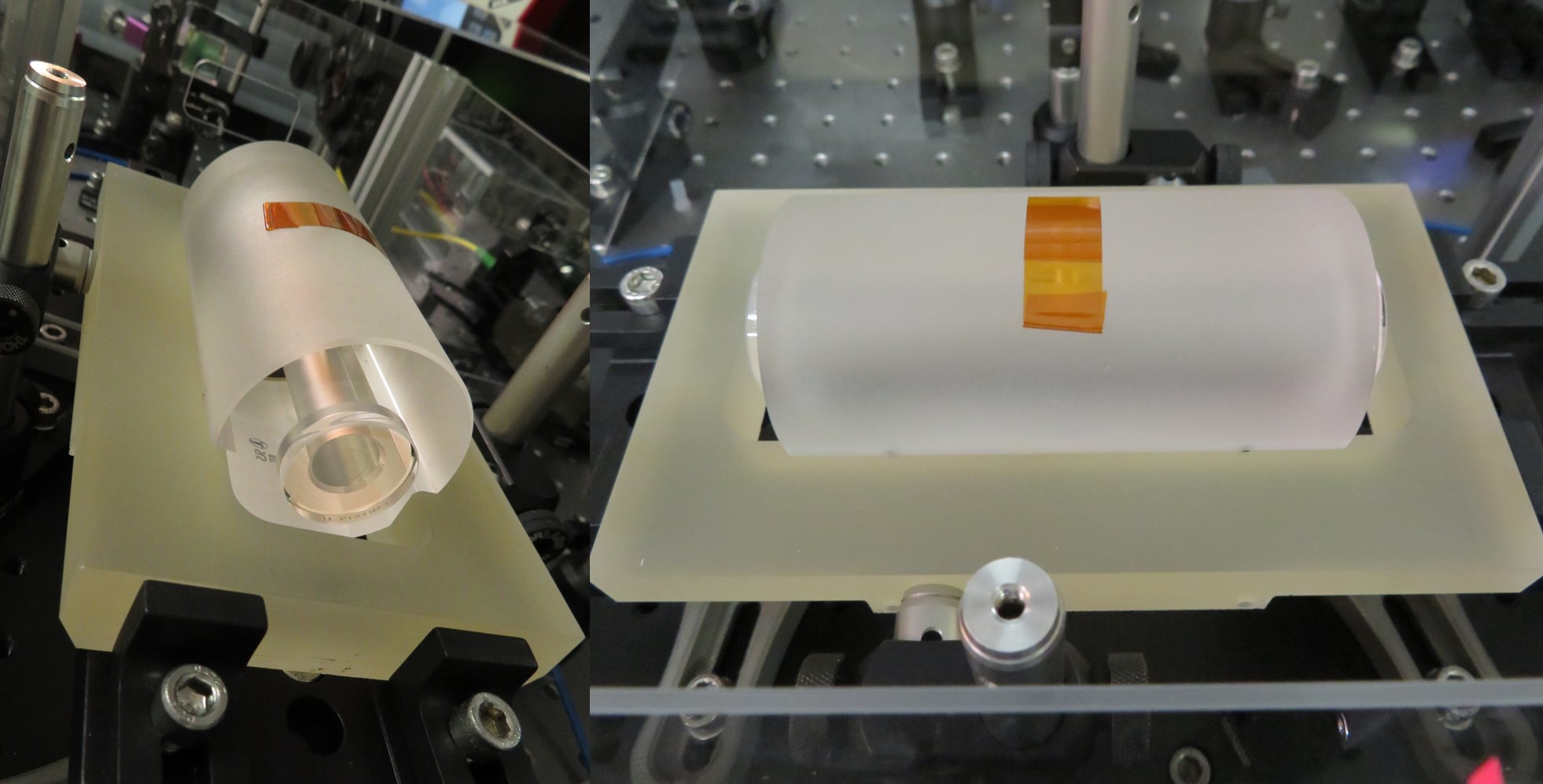}
\caption{Optical cavity resting on the carrier platform.}
\label{fig:cavityexp}
\end{figure}

The optical cavity used in my experiment is the plano-concave one. It has a diameter of 50 mm and length $L=10$ cm, thus a FSR of $1.4998\approx 1.5 $ GHz. The cavity was made by \href{http://www.stablelasers.com/fabry-perot-cavities/}{Stable Laser Systems (SLS)}. It is made from ultra-low expansion (ULE) glass such that it can be tuned to a temperature where in first order the length is stable to temperature fluctuations. The arrangement is shown in figure \ref{fig:cavityexp}. Ideally, the optical cavity must be placed in an isolated box with vacuum inside and a temperature control (for which I proposed a thermally controlled box with mechanical and thermal isolation).

After the transmission through the cavity is good and it is very well aligned (one way to check this is to have the maximum intensity for the TEM$_{00}$ mode and most of the other modes, if they exist, must have a centre symmetry). This can be achieved by finding a TEM$_{00}$ mode and scanning the frequency around this mode using the scan control. The a two mirror walk can help achieve the best alignment for the maximum transmission for this mode as seen on the oscilloscope.

\begin{figure}[hbtp]
\centering
\includegraphics[scale=0.34]{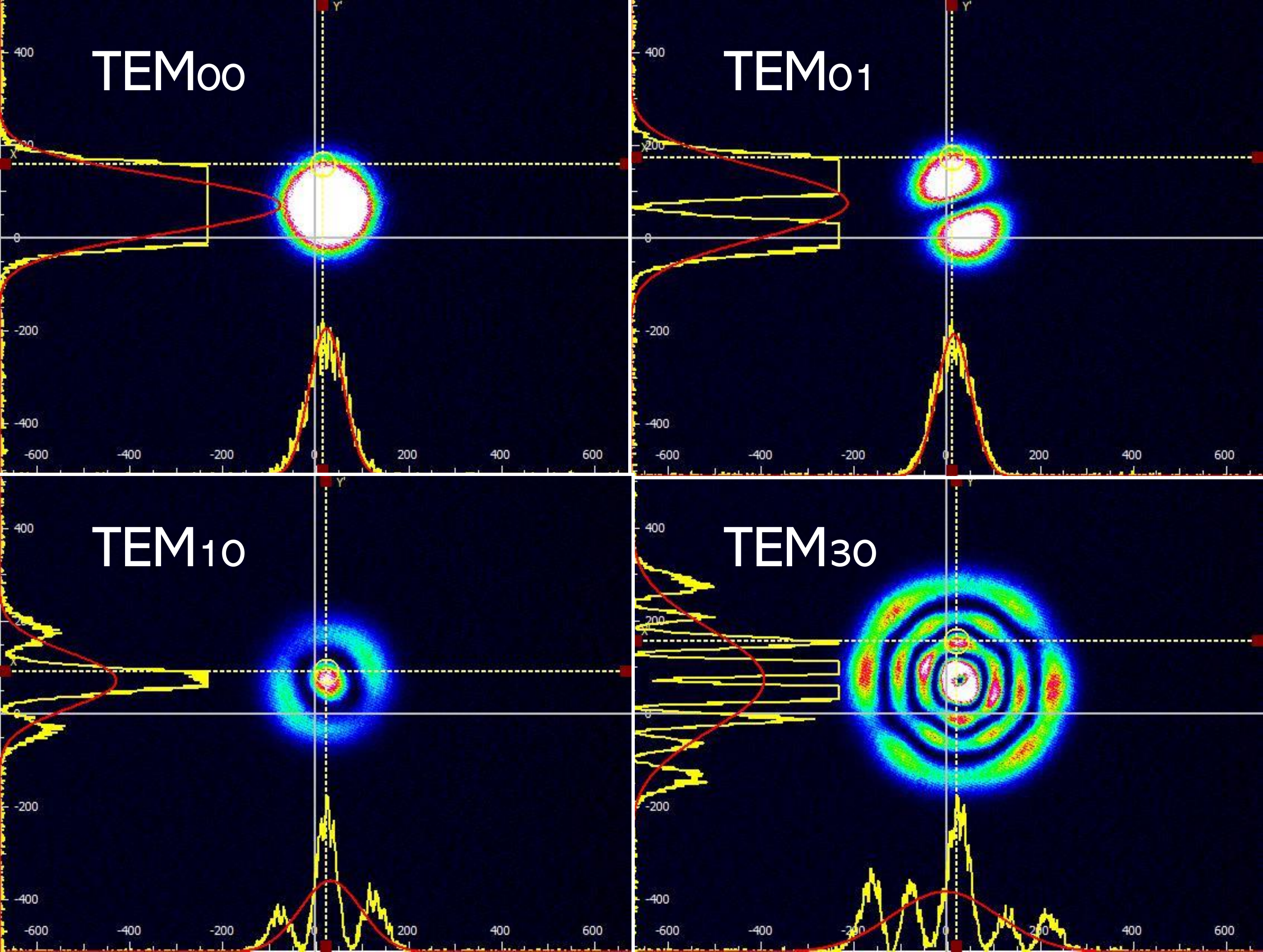}
\caption{Some frequently seen modes while aligning the optical cavity. These are the TEM$_{00}$, TEM$_{01}$, TEM$_{10}$ and TEM$_{30}$ respectively. Relative intensity profile is shown via the \href{https://www.thorlabs.com/newgrouppage9.cfm?objectgroup_id=3483}{ThorLabs Beam Profiler}. The conventional color scheme is used - Blue (low) to White (high). }
\label{fig:TEM}
\end{figure}
Various optical cavity laser modes obtained at different frequencies are shown in figure \ref{fig:TEM}. These are 2D intensity projections over the x-y plane. Axes profiles are also shown along with the centroid and the point of highest intensity.

The next step is to obtain the PDH signal. As already explained in the previous subsection \ref{subsec:EOM} regarding the detectors, the reflected signal (sidebands) are incident on another photodetector which send the electronic signal to the RF-lock box which after a proper power and phase shift selection output a desired waveform for the PDH signal. There are three parameters to take care of and optimize, to get the best PDH signal for the PID input. These are: the sideband power, the phase of the mixer and the power of the phase (the weight of the phase amplitude). These setting change all the time and new values can be quickly setup when the signal is observed on the scope. For the best readout of the error signal, the error signal must be optimized to have the maximum slope around the main carrier frequency and also have as long linear error waveform as possible with the help of these parameters.

\begin{figure}[hbtp]
\centering
\includegraphics[scale=0.34]{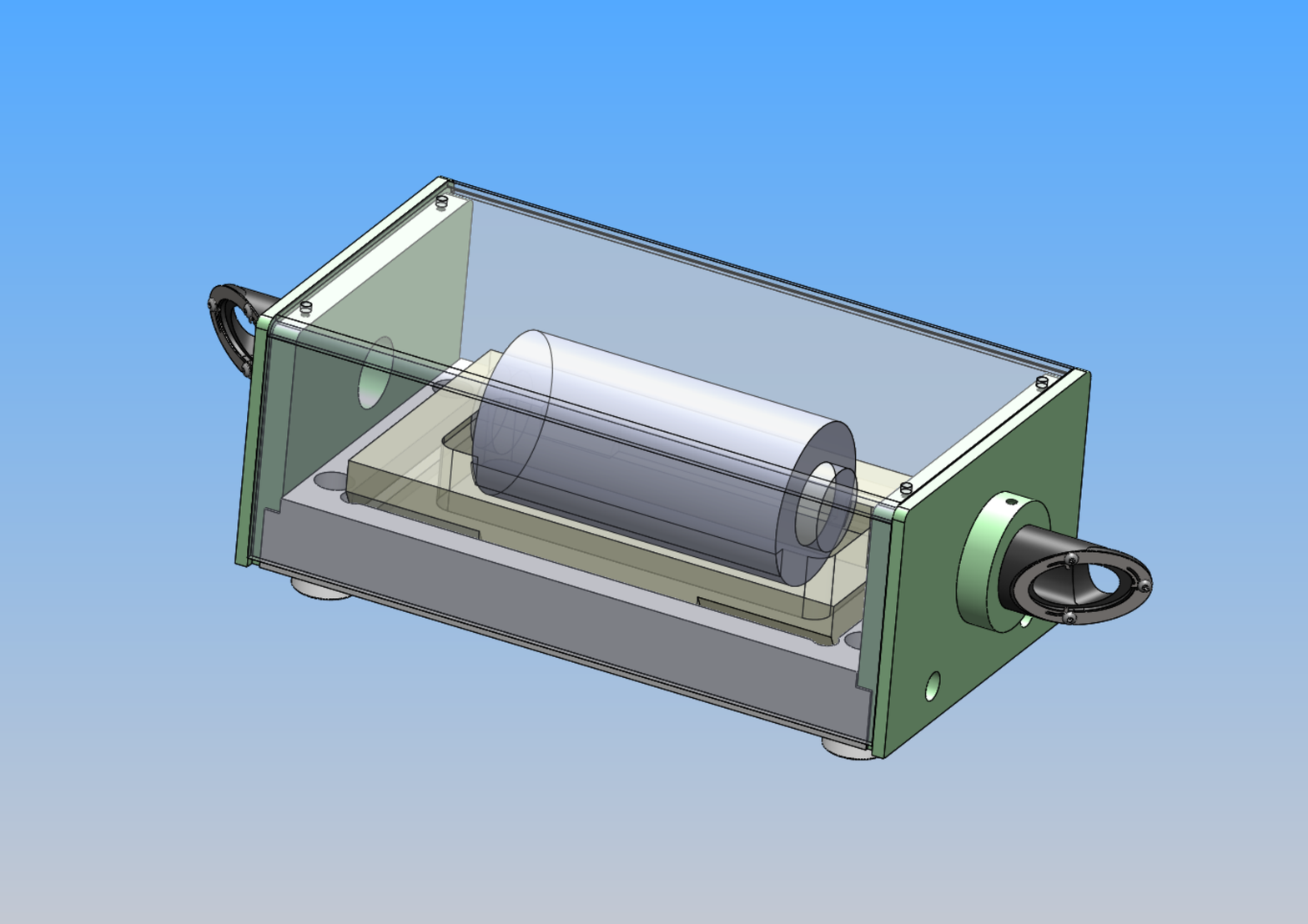}
\caption{Inventor design for the self designed cavity isolation box which is thermally controlled. \textit{Digital design and manufacturing credits: G. Martin, Universit{\"a}t Basel}.}
\label{fig:CavityBox}
\end{figure}

Another crucial part to look at is the \textit{noise}. First step is to figure out where the noise, that gives the unwanted drift to the laser frequency, originates from. It could be either the laser itself or also equally the optical cavity noise. A quick way to check it is to try locking with an isolated cavity and a noisy unprotected cavity. For my case, there was a lot of thermal and mechanical vibration noise to be eliminated. For this purpose, there are also some temperature controlled isolation boxes to place the cavity. It generally makes a lot of difference in the cavity performance as I show in the next subsection. The cavity has a special thermal point of temperature at which it does not expand or contract (hence no chance in length). This point lies around 30$^\circ$ C to 40$^\circ$ C for the cavity I am using. A thermal noise test on the cavity needs to be performed in order to determine this temperature precisely. For this purpose, I also designed a thermal isolation box shown in figure \ref{fig:CavityBox}. This is a aluminium box with the Brewster windows for the ideal transmission of laser light to the cavity. The box is heated with the a heating tape to the walls. A desired temperature maybe achieved and further thermal tests can be performed. 

\subsection{Optical cavity linewidth and finesse measurements}
In this section I will calculate some parameters which characterize the optical cavity mainly - linewidth and the finesse. In principle, if the length of the cavity and finesse/linewidth is known, everything else can be determined. 

The datasheet from the manufacturer for the cavity indicates that the cavity has a transmittance of 0.21380\% at $\approx 866$ nm wavelength. This gives a reflectance of 99.7862\%. Using equation (\ref{eq:finesse}), the ideal (reported) value of finesse for this cavity should be $\mathscr{F}=1467.8344$. Now using the finesse and FSR, we can estimate the linewidth using equation (\ref{eq:finlin}). This comes out to be $\nu_{1/2}=1.0212$ MHz.

Now I will describe and implement two methods for the estimation of linewidth and finesse experimentally.

\subsubsection{Linewidth measurement}
First of all, the the corresponding change in frequency with the piezo voltage must be calculated. It can easily be calculated from the oscilloscope or a voltmeter. For the LV mode of the controller box, 1 complete turn on the piezo voltage (= 2.4 V) corresponds to 430 MHz on the wavelengthmeter. Hence,
\begin{equation}
    1 \text{ V} \equiv 179.167 \text{ MHz}.
\end{equation}

\begin{figure}[hbtp]
\centering
\includegraphics[scale=0.42]{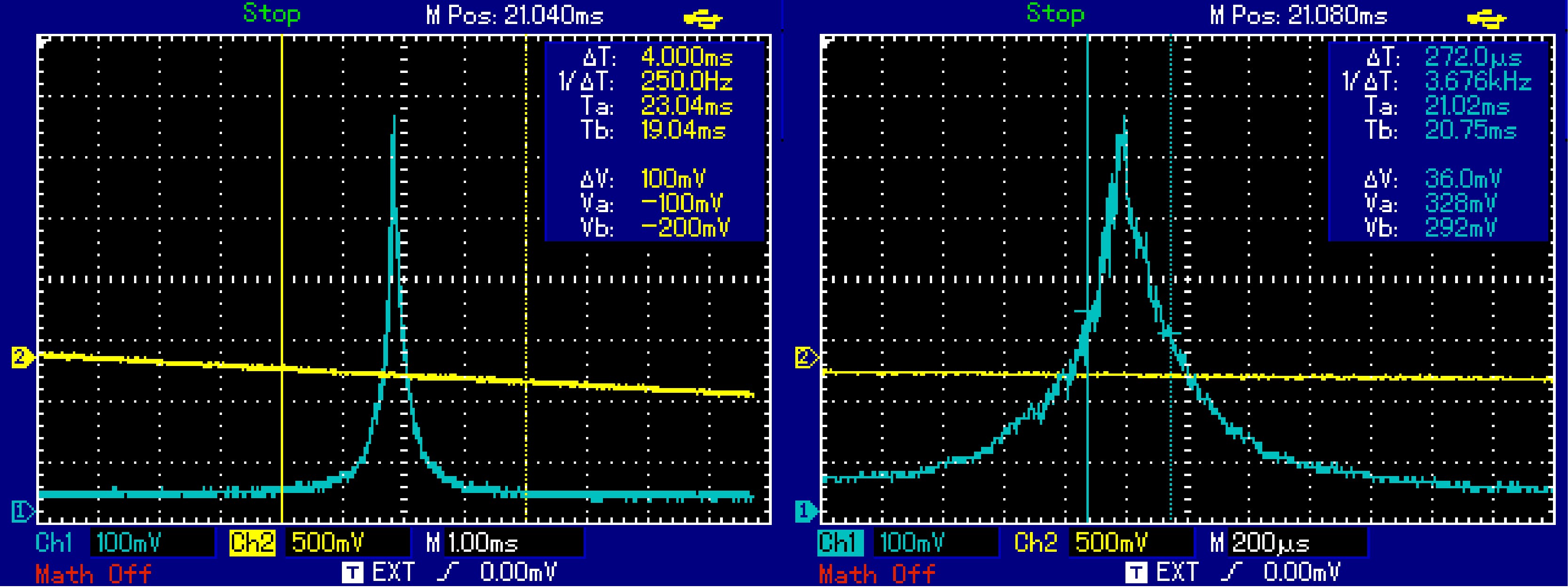}
\caption{The cavity transmission signal (blue) and the piezo ramp (yellow) when the cavity is scanned using the piezo about the TEM$_{00}$ mode. All other excited modes are highly suppressed in intensity when the fundamental mode is maximized.}
\label{fig:linewidth}
\end{figure}

Now, all measurements in voltages on the scope can be directly converted to frequency changes. Next step is to trigger the scope with the piezo scan signal and simultaneously look at the piezo ramp and the cavity transmission on the scope as shown in figure \ref{fig:linewidth}. Set the cavity scan amplitude such that the fundamental mode transmission peak lies completely on one of the ramps of the scan signal. This gives simultaneous measurement of the linewidth and the scan signal slope on the oscilloscope. One should make sure that the fundamental mode is maximized with the cavity alignment.

The FWHM (Full Width Half-Maximum) of the transmission peak can now be calculated using the scope cursors as shown. The measured linewidth in time scale can be directly converted to the voltage scale using the ramp slope. This can further be converted to the frequency scale using the previous equivalence. A repeated measurement gives a rough estimation of the linewidth and thus, finesse (because FSR is given).

The mean value of the linewidth from the scope, based on 20 independent measurements, is 1.0909 MHz, this gives the finesse value of $\mathscr{F}=1375.9796$ which is not very far from the given value. The linewidth measurement from the Digital Signal Oscilloscope (DSO) is not very accurate and precise. The signal does not have a very ideal waveform and the resolution is limited. However, as a means of measurement, is very helpful to classify the quality of the cavity.

\subsubsection{Cavity ring-down method for finesse measurement}
Cavity ring-down spectroscopy consists of determining the mean internal reflectivity $r$ of the lenses in an optical cavity by very quickly shut the light source and then measuring the time constant $\tau$ of the rate of decay of light from the cavity \cite{ref:20}. The decay rate of light intensity from the cavity takes the form of exponential decay as such: 
\begin{equation}
    I(t)=I_0e^{-t/\tau}
\end{equation}
And the relationship between $r$ (reflectivity) and $\tau$ is given by
\begin{equation}
    \tau=\frac{n}{c}\frac{l}{1-r+X}
    \label{eqn:decay}
\end{equation}
for $n$ is the index of refraction within the cavity (in my case $n$ = 1), $c$ is the speed of light, $l$ is the length of the cavity and $X$ is a constant accounting for optical losses due to
imperfections in the setup and photon absorption. In order to yield a usable decay curve, it is of paramount importance that the light shutoff time be significantly lower than the time
constant of the decay curve. This can be done electronically or even mechanically with a very quick shutter. Usually, an AOM is used for this purpose to quickly change the frequency off the resonance with a typical turn off time of 200 ns. Nearly the same can be achieved with a \textit{swift} (typically a GHz oscilloscope) oscilloscope when we can scan the frequency very fast. This is the approach I followed. For low loss limit, we can neglect $X$. The piezo was scanned such that the time scale of decay typically was observed in the range of 100 ns on the scope. For this case, equation (\ref{eqn:decay}) takes the form
\begin{equation}
    r=1-\frac{l}{c\tau}.
\end{equation}

\begin{figure}[hbtp]
\centering
\includegraphics[scale=0.28]{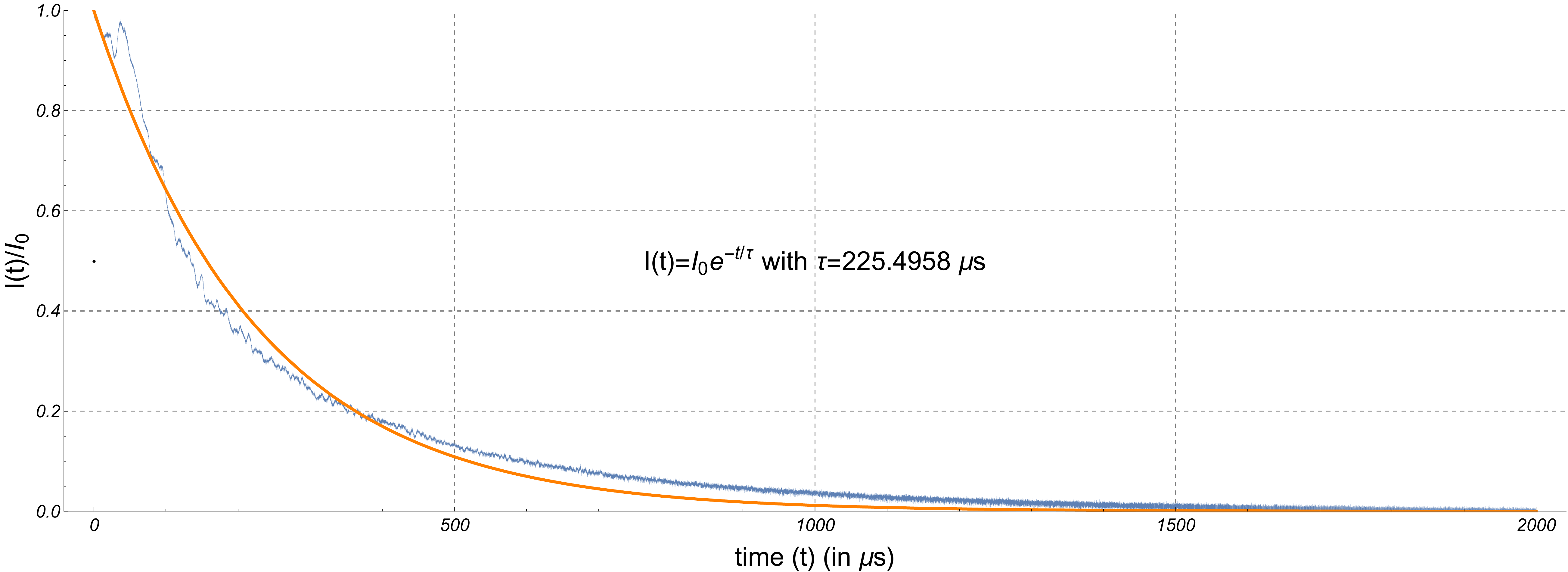}
\caption{The exponential decay (blue $\rightarrow$ experimental $\&$ orange $\rightarrow$ data fit) of the power in the cavity detected through the transmission photodetector.}
\label{fig:decay}
\end{figure}

Many such measurements were taken at different sweep (high) frequencies for the piezo. This gave the various estimates for $\tau$ value using which we can estimate the reflectivity of the mirrors. One such typical data set is shown in figure \ref{fig:decay} where the least square exponential fitting is done in Mathematica. The mean value of $\tau$ was found to be $\langle \tau \rangle=262.6692\text{ $\mu$s}$ which gives the reflectivity value of 0.998731 and finally a finesse value of $\mathscr{F}=1237.0167$. The linewidth can be conversely calculated using this value. Clearly, both the methods give close values for the finesse for the cavity.

\newpage
\section{Laser locking analysis}
After obtaining the best possible PDH signal, next step is its manipulation and feedback to the laser. The self-built 866 nm laser (figure \ref{fig:diodelaser}) has no modulation input on its own (some commercial lasers now have it). It is a simple ECDL whose current, piezo and temperature are externally modulated through the laser driver (figure \ref{fig:controller}) which has limitations for the modulation input bandwidth of its own as explained in the relevant section (\ref{subsec:controller}) earlier. 

\begin{figure}[hbtp]
\centering
\includegraphics[scale=0.5]{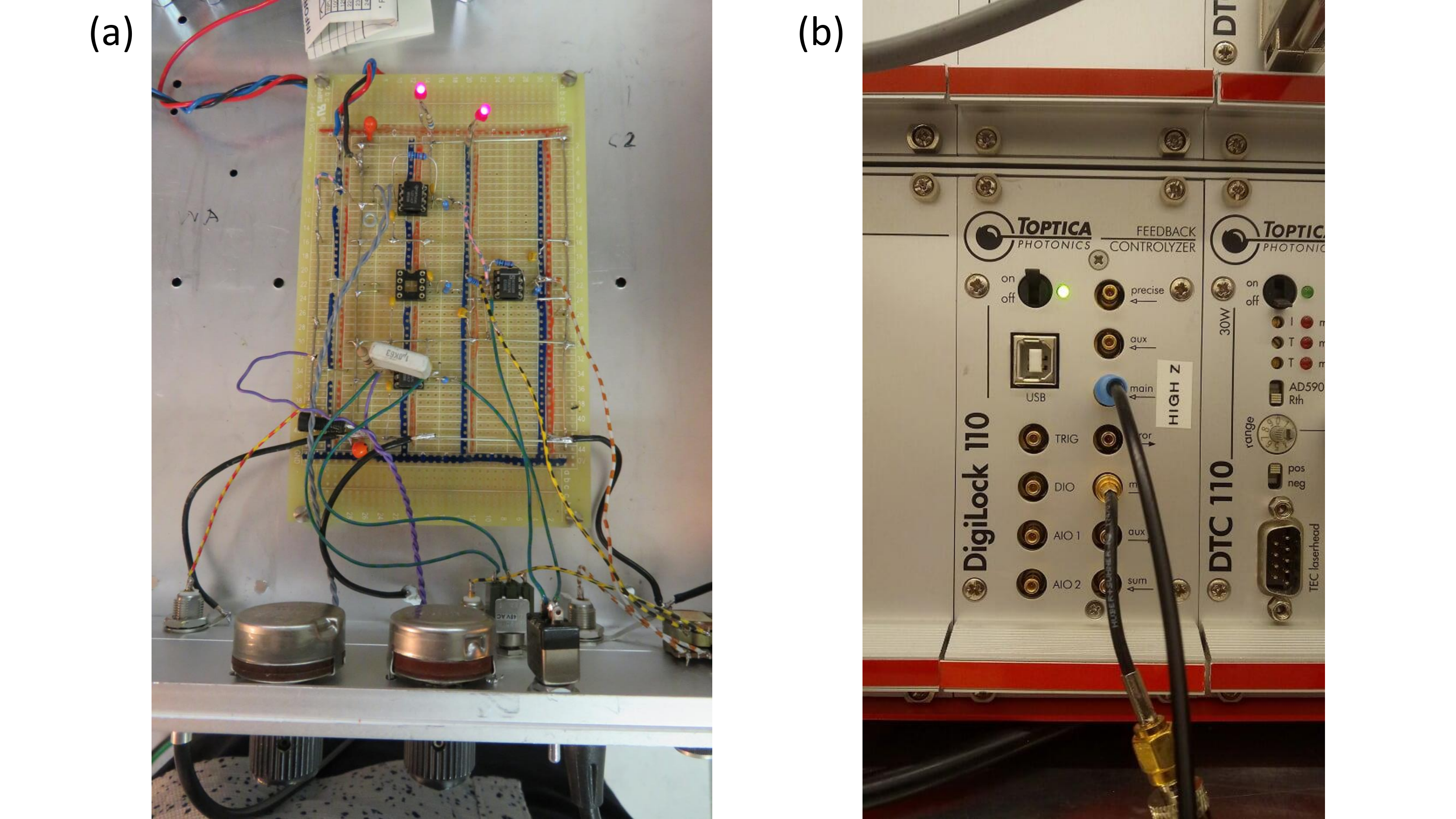}
\caption{A manually built PID controller with the \textbf{P} and \textbf{I} control output versus \href{https://www.toptica.com/fileadmin/Editors_English/03_products/03_tunable_diode_lasers/04_control_electronics/02_laser_locking_electronics/DigiLock_110/toptica_Digilock_Manual.pdf}{Toptica Digilock 110} module with dual PID outputs.}
\label{fig:PIDs}
\end{figure}

This PDH signal needs to be treated with a \textit{good} PID circuit whose output(s) will be fed to the modulation inputs of the laser controller. For the realization of this, I use two approaches. One is to use a self designed PID controller using the resistors, capacitors and ICs. Other, simply using a commercially available dual PID digilock - \href{https://www.toptica.com/fileadmin/Editors_English/03_products/03_tunable_diode_lasers/04_control_electronics/02_laser_locking_electronics/DigiLock_110/toptica_Digilock_Manual.pdf}{Toptica Digilock 110} (compatible with the SYS DC 110 controller I am using for my experiment). It is shown in figure \ref{fig:PIDs}.

While a home-built PID circuit can perform as well as any other PID or even better, it is much simpler and less time consuming to use a commercial locking box such as the Digilock 110. It comes with the digital interface which can be controlled from a computer and is much more adaptable even when the laser falls out of the lock, to restore it back. A specific point on the PDH curve slope can also be selected to lock the laser there. However, I have worked with both the systems, as the purpose of this work was to learn the experimental methods and pay more focus on 'building' the system and electronics from the scratch and hence learn the technique.

\subsection{Performance with the manually built PID circuit}
Now I present the performance of the manually built PID circuit. 
\begin{figure}[hbtp]
\centering
\includegraphics[scale=0.28]{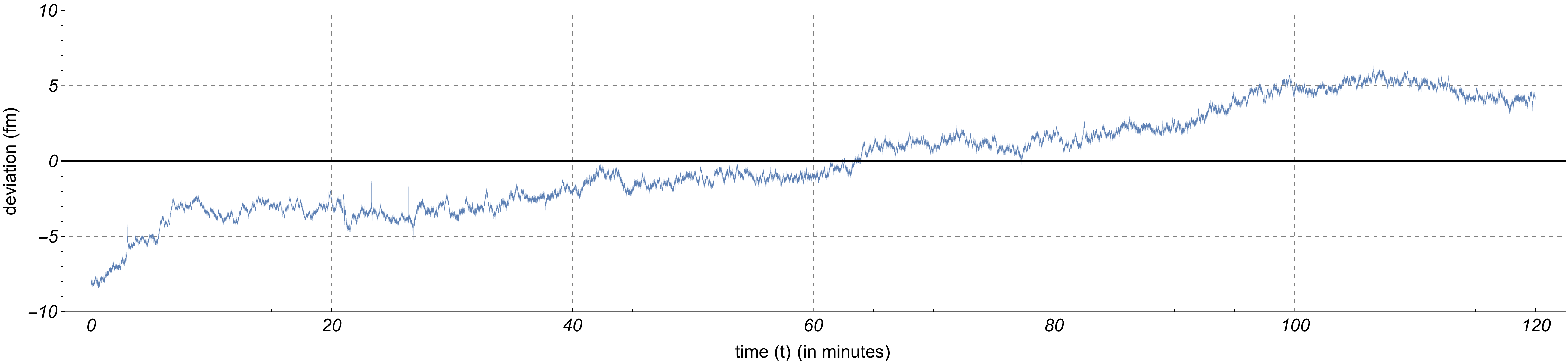}
\caption{Laser wavelength deviation of the unlocked laser with respect to time when for a duration of about 2 hrs. In this case, laser always continues to drift in one direction over a long scale of time.}
\label{fig:UnlockedWithLid}
\end{figure}
Figure \ref{fig:UnlockedWithLid} shows the time-deviation profile of the unlocked laser which is just isolated with a plastic sheet (low mechanical isolation). This is a comparison of the laser frequency to the wavelengthmeter (WM) frequency. The wavelength is measured with a wavelengthmeter which has a least count of 0.5 fm and the response rate of 1 ms for these measurements. It only measures the relative drifts between them. This laser shows an overall deviation of about 15 fm as a constant drift over a long time as compared to a wavemeter. One should note that the wavemeter doesn't measure an absolute frequency and the drifts are the sum of the cavity drifts which the laser in referenced to and the wavemeter drifts. Here, we assume that the short-term wavemeter fluctuations are $<0.5$ fm and also that the long-term drifts are mostly dominated by the cavity thermal expansion. The laser also has small scale fluctuations which are also of the scale of 5 fm.

I tried laser locking in mainly two categories. One is locking with the piezo and other locking with current control. Piezo locking is limited by the mechanical limitations of the piezo control. It is comparatively very slow responsive to the current modulation. It also has a low bandwidth. Current modulation is very speedy and has a very high bandwidth. Ideally, these should be modulated simultaneously using two separate PIDs which are independently controlled. The current modulation takes care of the small scale fluctuations and the sudden noise impulses that the setup faces while the piezo takes care of the long term drift of the laser. The piezo can change the wavelength over a long scale of about 0.1 nm. So, even if the laser goes badly out of the lock, it can be brought back using the piezo manually (which is not in the domain of current).

\subsubsection{Locking with Current only}
\begin{figure}[hbtp]
\centering
\includegraphics[scale=0.28]{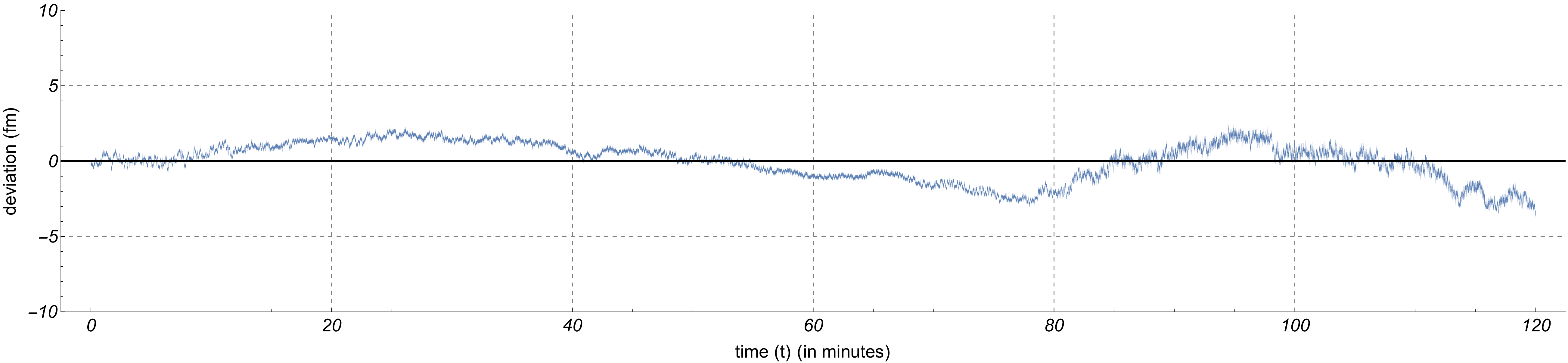}
\caption{Laser wavelength deviation when it is locked with current and is also thermally isolated but mechanically isolated. This is the best locking with the manual PID circuit. This is the best locking with the manual PID circuit. However, near the 80 minute, the laser deteriorates certainly due to a bad laser mode.}
\label{fig:LockedCurrentLidOn}
\end{figure}

\begin{figure}[hbtp]
\centering
\includegraphics[scale=0.28]{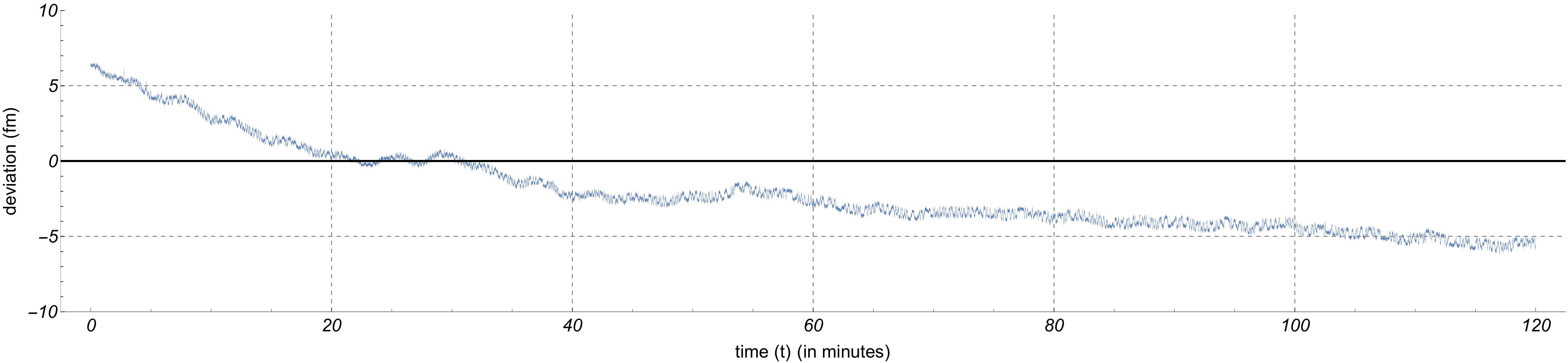}
\caption{Laser wavelength deviation when it is locked with current and is neither thermally isolated nor mechanically isolated. The laser continues to drift in one direction but with smaller fluctuations.}
\label{fig:LockedCurrentLidOff}
\end{figure}

Now, I present the locking with the current. The best locking obtained is shown in figure \ref{fig:LockedCurrentLidOn} with the plastic lid covered. The maximum deviation is only about $\pm 2$ fm which is very good over the unlocked laser for a duration of 2 hrs. This is already a good improvement over the unlocked laser. Now to further make the prediction about the nature of the large scale drift noise, I remove the plastic covering and let the cavity sit in the bare environment and subjected to all the local mechanical noise. I try to keep rest of the conditions as same as possible. This case is shown in figure \ref{fig:LockedCurrentLidOff}. This case is even much worse (overall deviation of about 30 fm) than an unlocked laser when the reference cavity is roughly isolated. This clearly suggests that the main noise is coming from the bad cavity attributes with the environment. The small scale deviations are still taken care with the current modulation, but the large scale drift is more enhanced. This may come from cavity length changes due to the temperature which now easily affects the cavity much more swiftly than before.

One possible origin of this large scale noise could be the length changes  in the cavity such that the resonance frequency of the cavity changes over the coarse of time. This could be due to the mainly the temperature changes from day/night which are observable enough for the timescale of few hours. Thus, it is very necessary to thermally isolate the optical cavity at the right temperature at which it has a stationary point for its thermal expansion coefficient. This motivated the construction of such a isolation box, in figure \ref{fig:CavityBox}. Just as a quick check, even waving hands near the cavity without the lid produced deviation of about 20 fm which is not a good sign. The laser can be very ideally locked when aided by such a good isolation as proposed.

\subsubsection{Locking with Piezo only}
\begin{figure}[hbtp]
\centering
\includegraphics[scale=0.28]{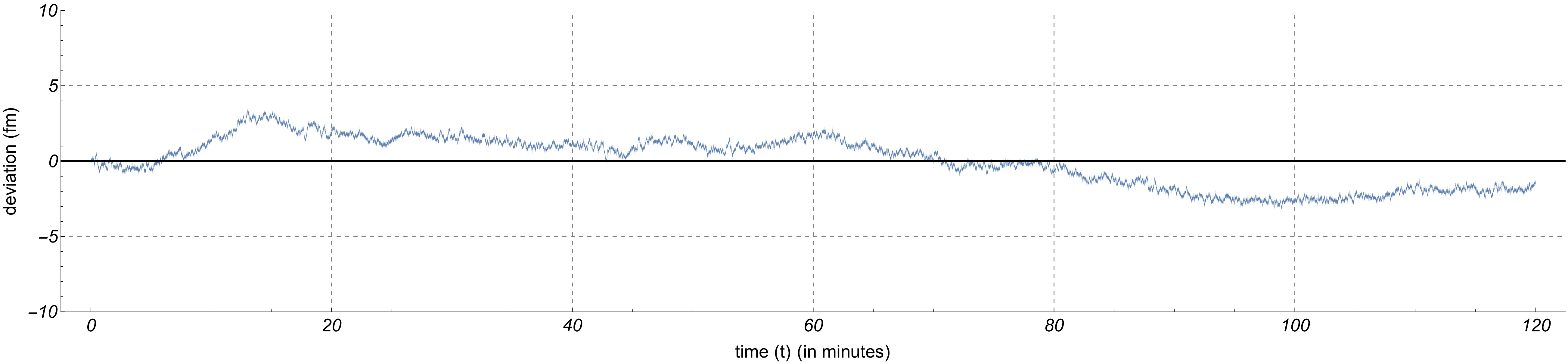}
\caption{Laser wavelength deviation when it is locked with piezo and is \textit{not} thermally isolated but is mechanically isolated.}
\label{fig:LockedPiezoLidOn}
\end{figure}

Similar locking is done using the piezo scan control. The plot for this is shown in figure \ref{fig:LockedPiezoLidOn}. This is also a good improvement over the unlocked laser but is not very sensitive for the low scale fluctuations. This can be compared with figure \ref{fig:LockedCurrentLidOn} or figure \ref{fig:LockedCurrentLidOff} which both show comparatively similar performance. For the best performance, a good cavity isolation and two independent PIDs are required to modulation both parameters (current and piezo modulation). Locking with a self-built PID circuit is not a very bad deal at the end. If two PIDs are separately constructed, they can do the locking task (stable over at least the scale of few hours, laser does not fall out of the lock over a coarse of just few hours), in practice. 

\begin{figure}[hbtp]
\centering
\includegraphics[scale=0.28]{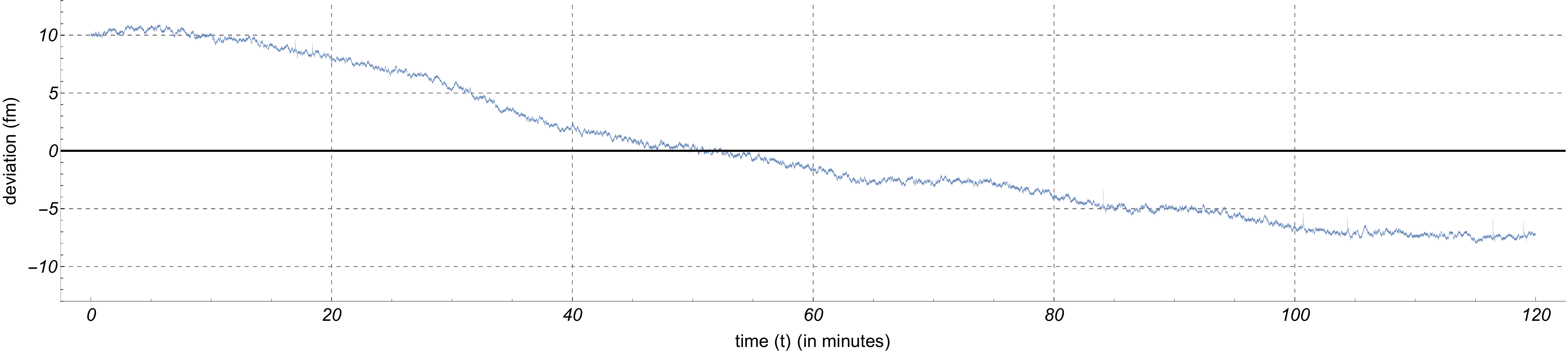}
\caption{Laser wavelength deviation when it is locked with current and is neither thermally isolated nor mechanically isolated.  The laser continues to drift in one direction.}
\label{fig:LockedPiezoLidOff}
\end{figure}

A similar behavior is shown as in figure \ref{fig:LockedCurrentLidOff} when the mechanical isolation (the plastic box) is removed. The laser continues to drift in one direction over the coarse of time. This is plotted in figure \ref{fig:LockedPiezoLidOff}.

\subsection{Performance with the Digilock circuit}
When an advanced feedback system, like the Digilock, was used, it significantly improved the system stability. The laser drifted only about $\pm1.5$ fm over the same scale of time. 
\begin{figure}[hbtp]
\centering
\includegraphics[scale=0.28]{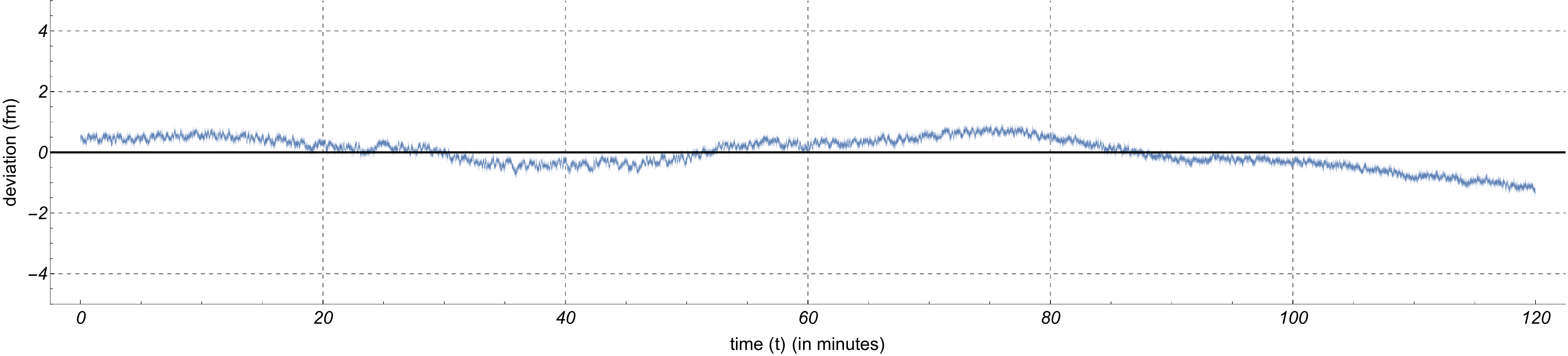}
\caption{Laser wavelength deviation when it is locked with with the dual PID Digilock and the cavity is mechanically isolated (not thermally). Two PIDs are simultaneously used to feed both, the current and the peizo modulation. This is significant improvement to the system.}
\end{figure}

However, it should be kept in mind that the above data is taken considering the wavelengthmeter to be ideal, which is not the case. The wavelengthmeter has its own wavelength drift which introduces error in the measurements.

\subsection{Remarks on the laser locking}

From the above results, we conclude that a laser stabilization system can be realized with the PDH technique using simple optics and electronics. The system works very well but it is yet somewhat far from perfection which can be achieved, in principle. There are some limitations of the current system:
\begin{itemize}
    \item \textbf{Slow response of the laser}: Any ECDL, commercial or home built, features high frequency noise which we couldn't compensate due to the low bandwidth of our electronics. This means that the laser linewidth or the short terms fluctuations could not be completely suppressed. We could see this directly by the fact that our transmission signal from the cavity was fluctuating a lot, a signal that the laser drifts were larger than the cavity linewidth. This problem can still be tackled using a fast responsive mediator or building an electronics board which can increase the laser response rate. Such a board acts like a bridge between the laser controller and the laser itself so that the laser can respond to a wide range of feedback signals  \cite{ref:14}.
    \item \textbf{No thermal isolation}: As I pointed out before, the cavity must be stored in a thermally controlled environment to prevent it from capturing the environmental thermal changes \cite{ref:15}. The temperature must be finely controlled. The cavity has a very special characteristic that in the neighbourhood of a certain temperature, it does not expand or contract with small temperature fluctuations. This temperature maybe found precisely using independent measurements in the cavity.
    \item \textbf{No ideal mechanical isolation}: This part can still be improved by using better optical platform for the cavity placement. The current mechanical isolation is simply a plastic cover which only prevents the local air motion from affecting the cavity. A good mechanical isolation scheme would prevent it even from sudden impulses which are very frequent \cite{ref:17} \cite{ref:18}. Ideally, both of the above problems can be resolved by placing the cavity in a thermal vacuum shield which is stored in a mechanically isolated box which prevents all reasonable vibrations. Such a box could be heated through radiation to attain a certain temperature. Stable cavity housings are usually similar to the ones given \href{http://www.stablelasers.com/cavity-housings/}{here}.
    \item \textbf{Instability of the optical cavity}: A high finesse cavity (typically $\mathscr{F}>100000$) with the smallest linewidth possible maybe better to use. Fine wavelength selection for such a cavity will be very precise. If the linewidth of the cavity itself is large, it will transmit the neighboring frequencies along with the resonant frequency which would not count as a the laser drift \cite{ref:19}. However, high-finesse cavities are much harder to lock.  In our case the D$_{3/2}$ ↔ P$_{1/2}$ transition in $^{40}$Ca$^+$ has a linewidth of $\sim$1 MHz. With a cavity with linewidth of 1 MHz one can hope well to lock the laser below 100 kHz which is what necessary for these transitions.
\end{itemize} 

\section{Conclusion}
In this work, I described the theory for Pound-Drever-Hall method for laser stabilization and how one can built the stabilization system. All theoretical concerns of the technique such as noise and stability are addressed. Cavity characteristics are also estimated through two methods of finesse estimation. The main objective of the work was to built the Pound-Drever-Hall laser stabilization system from scratch and deal with the necessary improvements over the noise. All the major concerns in relation to to the PDH method are addressed. It was inferred that thermal and mechanical isolation of the optical cavity play a major role for stability, and they can be minimized by relatively straightforward methods as stated.

\textbf{\textit{Acknowledgements}} - I am very thankful to Prof. Stefan Willitsch for giving me this opportunity to spend a summer at his Quantum Technology (QuTe) and Ion-Trap lab which made the experimental internship possible and IAESTE Switzerland for assisting with the visit. I would also like to thank Dr. Ziv Meir for his constant guidance throughout the internship and Mr. Mudit Sinhal for valuable discussions in times of need. I am very thankful to Ms. Priya Lakshmi for introducing me to various methods in experimental optics.

\newpage
\bibliographystyle{unsrt}
\bibliography{Bibliography.bib}


\end{document}